\documentclass{aa}

\usepackage{graphicx}
\usepackage{txfonts}
\usepackage{hyperref}

\usepackage{amsmath,amssymb,bm,bbm}
\usepackage{physics}

\newcommand\ii{{i}}

\newcommand\id{{\mathrm d}}

\newcommand\cL{{\mathcal L}}

\newcommand\br{{\bm r}}

\newcommand\bg{{\bm g}}

\newcommand\bu{{\bm u}}

\newcommand\bxi{{\bm \xi}}

\newcommand\Cref{C_{\text{ref}}}
\newcommand\cref{\Cref}

\newcommand\bPsi{{\bm \Psi}}
\newcommand\unitr{\bm{\hat{r}}}
\newcommand\unitth{\bm{\hat{\theta}}}
\newcommand\unitph{\bm{\hat{\phi}}}

\newcommand\pOneg{p_{1,g}}
\newcommand\rhoOneg{\rho_{1,g}}
\newcommand\PhiOneg{\Phi_{1,g}}
\newcommand\pOnegt{\tilde{p}_{1,g}}
\newcommand\rhoOnegt{\tilde{\rho}_{1,g}}
\newcommand\PhiOnegt{\tilde{\Phi}_{1,g}}

\newcommand\cSqOnegt{\tilde{c}^2_{1,g}}

\newcommand\bgOneg{\bg_{1,g}}

\newcommand\divxip{\mathrm{Div}_{\bxi_p}}

\newcommand\GYRE{\emph{GYRE}}
\newcommand\ADIPLS{\emph{ADIPLS}}
\newcommand\MESA{\emph{MESA}}

\usepackage{xcolor}

\usepackage{CJKutf8} %Chinese typesetting package

\begin{document} 
\begin{CJK*}{UTF8}{gkai}

   \title{Signature of solar g modes in first-order 
   p-mode frequency shifts}

\author{
	Vincent G. A. Böning\inst{1}
	\and
	Huanchen Hu\inst{1,2}(胡奂晨)
	\and
	Laurent Gizon\inst{1,3,4}
}

\institute{
	Max-Planck-Institut f\"ur Sonnensystemforschung, Justus-von-Liebig-Weg 3, 37077 G\"ottingen, Germany \\
	\email{boening@mps.mpg.de}
	\and
	{Max-Planck-Institut f\"ur Radioastronomie, Auf dem H\"ugel 69, 53121 Bonn, Germany}
	\and
	Institut für Astrophysik, Georg-August-Universität G\"ottingen, Friedrich-Hund-Platz 1, 37077 Göttingen, Germany
	\and
	Center for Space Science, NYUAD Institute, New York University Abu Dhabi, PO Box 129188, Abu Dhabi, UAE
	}

\date{Received 8 March 2019 / Accepted 27 June 2019}

% \abstract{}{}{}{}{} 
% 5 {} token are mandatory

\abstract
{Solar gravity modes (g modes) are buoyancy waves that are trapped in the solar radiative zone and have been very difficult to detect at the surface. Solar g modes would complement solar pressure modes (p modes) in probing the central regions of the Sun, for example the rotation rate of the core.}
{A detection of g modes using changes in the large frequency separation of p modes has recently been reported. However, it is unclear how p and g modes interact. The aim of this study is to evaluate to what extent g modes can perturb the frequencies of p modes.}
{We computed the first-order perturbation to global p-mode frequencies due to a flow field and perturbations to solar structure (e.g. density and sound speed) caused by a g mode. We focused on long-period g modes and assumed that the g-mode perturbations are constant in time. The surface amplitude of g modes is assumed to be $1$~mm~s$^{-1}$, which is close to the observational limit set by Doppler observations.}
{Gravity modes do perturb p-mode frequencies to first order if the harmonic degree of the g mode is even and if its azimuthal order is zero. The effect is extremely small. For dipole and quadrupole p modes, all frequency shifts are smaller than $0.1$~nHz, or $2\times10^{-8}$ in relative numbers. This is because the relative perturbation to solar structure quantities caused by a g mode of realistic amplitude is of the order of $10^{-6}$ to $10^{-5}$. Additionally, we find that structural changes dominate over advection. Surprisingly, the interaction of g and p modes takes place to a large part near the surface, where p modes spend most of their propagation times and g modes generate the largest relative changes to solar structure. This is due to the steep density stratification, which compensates the evanescent behaviour of g modes in the convection zone.}
{It appears to be impossible to detect g modes solely through their signature in p-mode frequency shifts. Whether g modes leave a detectable signature in p-mode travel times under a given observational setup remains an open question.}


\maketitle
\end{CJK*}

%
%-------------------------------------------------------------------
%-------------------------------------------------------------------

\section{Introduction}

Using observations of pressure modes (p modes), helioseismology has been very successful in determining solar interior structure and differential rotation in the convection zone and in most of the radiative zone \citep[see, e.g.][]{Howe2009,Aerts2010, Basu2016}. 
However, a number of outstanding questions about the deep solar interior remain, such as the rotation of the solar core.

Solar internal gravity modes (g modes), if detected, would provide complementary information to the p modes. This is because g modes have most of their kinetic energy in the central regions of the Sun. In addition, they have a smaller radial wavelength in this region \citep[e.g.][]{Berthomieu1990,Berthomieu1991}, which would give a higher spatial resolution than the p modes.

The quest for detecting solar g modes has, therefore, attracted considerable attention in the past with several reported detections \citep[see][for reviews]{Appourchaux2010,Appourchaux2013}. Unfortunately, none of the reported detections have been confirmed so far.

In line with \citet{Kennedy1993} and \citet{Duvall2004}, \cite{Fossat2017} proposed a new method for detecting g modes  in the first-arrival travel time of low-degree p modes using disk-integrated observations. Under the simplest interpretation, this p-mode travel time is inversely proportional to the large frequency separation of p modes, which is the frequency difference between p modes with identical harmonic degrees whose radial orders differ by one. The p-mode travel time may be affected by a g-mode oscillation through changes in sound speed, for example. \cite{Fossat2017} measured the p-mode travel time using eight-hour intervals of Global Oscillations at Low Frequency \citep[GOLF,][]{Gabriel1995} data and a sampling rate of four hours. The idea is that such a time series can sample changes in solar structure and dynamics due to high-order g modes that have periods longer than eight hours. Using this method, \cite{Fossat2017} and \cite{Fossat2018} reported periodic signals in the power spectrum of the data that were interpreted in terms of the rotational splittings of the g modes, implying a solar core rotating nearly four times faster than the radiative zone. Subsequently, \cite{Schunker2018} reproduced the signal of \cite{Fossat2017}, but warned that it is very sensitive to a number of choices and parameters in the analysis procedure. Very recently, \citet{ScherrerGough2019} also showed that the statistical significance of the signal depends on the helioseismic instrument used (from SOHO, SDO, or from the ground).

The motivation for the present study is to understand to what extent g modes can affect p-mode frequencies and whether and how such an effect would be measurable. We do not address the question of the signature of g modes in the observed p-mode travel times, which will require a separate study in order to include the details of the measurement procedure.

One may first ask the general question whether p and g modes should interact at all. In a linear wave equation, the superposition principle holds. It ensures that any two wave fields can be added and still fulfil the wave equation. As a consequence, the presence of one wave does not change the properties of a second wave at all. The situation is different, however, when non-linear effects are taken into account. Non-linearities in the wave equation include non-linearities in the Navier-Stokes equations. For example, this is the case for advection, temporal changes in the background (sound speed, density, pressure), dissipation, and non-adiabatic effects \citep[e.g. Chapters~2 and~3 in][]{Hamilton1998}. In a non-linear wave equation, two solutions to the wave equation do not, in general, add up to a third solution because the waves interact. This effect may in principle be exploited. A successful example of detecting the presence of waves in the Sun using other waves  is given by the recent detection of solar equatorial Rossby waves, which were detected because they advect p modes \citep{Loeptien2018,Liang2019Rossby,Hanasoge2019}.

In classical acoustics, the non-linear scattering of sound by sound has been observed experimentally \citep[e.g.][]{Thuras1935} and modelled theoretically \citep[e.g.][and references therein]{Westervelt1957a,Westervelt1963,Hamilton1998}. For example, \cite{Westervelt1957a} obtained an approximative non-linear acoustic wave equation by inserting a solution to the linear wave equation into the full non-linear Navier-Stokes equations. In addition,  the non-linearity introduced by a second order term in the equation of state was taken into account. In the presence of sound waves of two different frequencies, the generation of waves with frequencies equal to the sum and difference of the original frequencies is observed \citep[e.g.][]{Thuras1935}.

The non-linear coupling of stellar oscillation eigenmodes has been studied by several authors. \cite{Perdang1982} used a Hamiltonian formalism and showed that a non-linear coupling can lead to chaotic motions with an approximate periodicity. 
\citet{Dziembowski1982} studied non-linear interactions of two and three modes and found that two-mode interaction has an amplitude-limiting effect on stellar pulsations \citep[see also][]{Kumar1989}. Based on \citet{Dziembowski1982}, \citet{Wentzel1987} showed that two p modes can generate a g mode by a non-linear interaction. 
\citet{Wentzel1987} argued that the evanescent behaviour of the g modes in the convection zone is overcompensated by the increase of p-mode sensitivity towards the surface. As a consequence, the coupling would take place predominantly in the convection zone. The non-linear coupling of p and g modes was also studied for tidally interacting neutron star binaries by \citet{Weinberg2013}, where the coupling is found to alter the phase of the emitted gravitational wave and thereby its detection signal. \citet{Dziembowski1982} considered modes that are in a resonance condition, that is one frequency is the sum or difference of the other two frequencies. In \citet{Weinberg2013}, the mode frequencies are non-resonant, but the radial wave numbers of the p and g modes are similar in a region in the stellar interior, which means that the radial wave numbers fulfil a resonance condition.

A possible signature of g modes in the p-mode frequency signal has been studied previously by \citet{Kennedy1993}, \citet{Lou2001}, and \cite{ScherrerGough2019}. In these studies, the g-mode period was assumed to be much longer than the p-mode period, and hence the perturbation to solar structure caused by a g mode was assumed to be time-independent on shorter temporal scales. \citet{Kennedy1993} used perturbation theory on the asymptotic p-mode dispersion relation and found that g modes can produce a frequency modulation of the p modes. This would create sidelobes near each p-mode frequency, which was also found by \citet{Lou2001}. Using degenerate perturbation theory, \cite{ScherrerGough2019} gave an expression for the first-order perturbation of p modes by g modes and estimated the order of magnitude of the signal. They found that the signal reported by \citet{Fossat2018} would be due to relative frequency shifts of the order of $10^{-5}$, which would correspond to a g-mode amplitude of $50\,\rm{cm}\,\rm{s}^{-1}$ and which is much higher than the observational upper limit on g-mode amplitudes \citep[for a review, see][]{Appourchaux2010}. Both \citet{Kennedy1993} and \cite{ScherrerGough2019} found that only g modes with even harmonic degrees can couple to p modes to leading order.

As the non-linearity in the wave equation is certainly a small effect, the interaction of p modes with g modes can be studied using perturbation theory. The simplest approach is to model the effect of a time-independent perturbation on mode frequencies \citep[][]{Kennedy1993,ScherrerGough2019}. In addition, it is possible to consider the change in eigenfunctions \citep[e.g.][]{Lavely1992}, changes in mode amplitude ratios \citep[e.g.][]{Schad2011b,Schad2013}, or the coupling of oscillations at different frequencies \citep[e.g.][]{Woodard2007,Woodard2016,Hanasoge2017FlowSoundSp}.

In this work, we expand on the approach taken by \cite{Kennedy1993} and \citet{ScherrerGough2019}, and study the first-order perturbation of p-mode frequencies by a frozen g mode, that is we assume that the perturbation caused by the g mode is constant in time. Our aim is to obtain quantitative estimates of the induced frequency shifts, and to understand in a more detailed manner where the interaction is taking place and which of the perturbative effects of the g mode is causing the interaction.

The effect of aspherical perturbations on p-mode frequencies has been studied by \citet{Gough1993}, on which \citet{Kennedy1993} and \citet{ScherrerGough2019} are based. The aforementioned work was done under the assumption that the perturbation to solar structure preserves hydrostatic equilibrium. In this paper, we include the effect of a departure from hydrostatic equilibrium as it is the case for a g mode. In addition, we provide detailed quantitative estimates of the induced first-order frequency shift and we provide a detailed analysis of the resulting selection rules. 
To do that, we will include a number of effects such as changes to solar structure through changes in sound speed, pressure, density, gravitational potential, and advection due to the flow field from a g mode.

In addition, we need numerical solar p and g mode eigenfunctions which can be computed using stellar oscillation codes like \ADIPLS~\citep{JCDadipls} or \GYRE~\citep{Townsend2013}. Such a computation is based on a stellar model. In this work, we use model S \citep{JCD1996} and a result from the \MESA~stellar evolution code \citep{Paxton2011}.

This paper is structured as follows. We first introduce the oscillation equations (Section~\ref{secOscEq}) and the changes in structure and dynamics due to a g mode (Section~\ref{secgmodeef}). In Section~\ref{secperturbtheory} we setup the perturbation theory and we derive the equations for the first-order frequency shift of a p mode in the presence of a g mode{, including selection rules}.  We then present numerical results in Section~\ref{secResults} and conclusions in Section~\ref{secConclusion}.

\section{Linearised equations of stellar oscillations}
\label{secOscEq}

To model the effect of solar g on p modes, we briefly review the equations of stellar oscillations following \citet[][Chapter~3]{Aerts2010} and \cite{Basu2016}. Solar g and p modes are characterised by an eigenfunction $\bxi(\br)$ which describes the displacement of a parcel of gas during an oscillation, depending on the location $\br$ in the solar interior. The displacement eigenfunction is a solution to the linear wave equation
\begin{align}
\omega^2\bxi = \cL[\bxi] 
, \label{eqwave}
\end{align}
where $\omega$ is the cyclic eigenfrequency of the mode and the wave operator $\cL$ can be written
\begin{align}
\cL[\bxi] = \frac{1}{\rho_0} \left( \nabla p_1 - \rho_1\bg_0 - \rho_0\bg_1 \right)
 - 2 \ii \omega (\bu_0\cdot\nabla ) \bxi.
\label{eqwaveOp}
\end{align}
Here, $\rho$, $p$, and $\bg$ denote density, pressure, and gravitational acceleration, and $\bu_0$ is a background flow, which we assume to be equal to differential rotation \citep[e.g.][]{Schou1998a},
\begin{align}
    \bu_0 =  r \, \sin\theta \,\Omega(r,\theta) \,\unitph.
\end{align}
Quantities with a zero subscript are values in hydrostatic equilibrium of the Sun. Quantities with a one as subscript denote Eulerian first-order perturbations to hydrostatic equilibrium and their explicit dependence on the displacement is given in Appendix~\ref{appWaveOp}.

It can be shown \citep{Lynden1967} that the wave operator $\cL$ is self-adjoint {under} the scalar product
\begin{align}
	\langle \bxi,\bxi'\rangle &= \int_\sun \bxi^* \cdot \bxi' \rho \, \id^3 \br
\end{align}
for a suitable boundary condition, for example a vacuum boundary with $\nabla\cdot \bxi = 0$. 
As a consequence, two eigenfunctions $\bxi$ and $\bxi'$ with eigenvalues $\omega^2\neq\omega'^2$ are orthogonal with regard to the above scalar product. For comparison purposes, we use four different boundary conditions in this work, see Section~\ref{secgmodeef} and Appendix~\ref{appBC}.

In hydrostatic equilibrium {and in absence of a background flow ($\bu_0=0$)}, the Sun is assumed to be a spherically symmetric star. In this case, the eigenfunctions can be written in terms of spherical harmonics,
\begin{align}
\bxi_{lmn}(r, \theta, \phi) &= 
\xi_{r,ln}(r) Y_{lm}(\theta, \phi) \unitr +\xi_{h,ln}(r) \left(\frac{\partial Y_{lm}}{\partial \theta} \unitth + \frac{1}{\sin \theta}\frac{\partial Y_{lm}}{\partial \phi} \unitph\right) \label{eqbxi} \\
&=\xi_{r,ln}(r) Y_{lm}(\theta, \phi) \bm{\hat{r}} +\xi_{h,ln} \bPsi_{lm}, \label{eqxiYlm}
\end{align}
where $\bPsi_{lm}$ is a vector spherical harmonic \citep{Barrera1985}, the quantities $\unitr,\unitth,\unitph$ are unit vectors in the directions of $r,\theta,\phi$, the functions $\xi_r,\xi_h$ are radial and horizontal displacement eigenfunctions, and the quantum numbers $l,m,n$ denote harmonic degree, azimuthal order, and radial order, respectively. We adopt the usual convention that the radial order $n$ for a p mode is positive and for a g mode is negative \citep[e.g.][Chapter~3]{Aerts2010} and normalise the eigenfunctions so that
\begin{align}
    \langle \bxi_{lmn},\bxi_{l'm'n'} \rangle =\delta_{ll'}\delta_{mm'}\delta_{nn'}. \label{eqeforthonormal}
\end{align}

The linearity of the wave equation implies that any mode, for example a p mode, is not affected to first order by the restoring force of a different mode, for example a g mode. By first order, we mean here the linearization of the Navier-Stokes equations that yields the linear wave equation. However, as g modes do perturb the hydrostatic equilibrium of the Sun, they may have some affect on p modes at a higher order, if non-linear effects are taken into account.

\section{Perturbation to solar structure and flows due to g modes}
\label{secgmodeef}

In order to determine the effect of a g mode on p-mode frequencies, we need to determine the perturbation of a g mode to solar structure. To do so, we computed eigenfunctions for g and p modes using the codes \GYRE~\citep{Townsend2013,Townsend2018} and \ADIPLS~\citep{JCDadipls}. The computations are based on three non-rotating solar equilibrium models, namely solar model S \citep{JCD1996}, a version of model S where the density was smoothed slightly, and a solar model obtained from \MESA~\citep{Paxton2011}, see Appendix~\ref{appBC}. Furthermore, we tested the effect of several surface boundary conditions like zero Lagrangian pressure perturbation (vacuum boundary) at the surface and two conditions for an isothermal atmosphere \citep{JCDadipls,Unno1989}, as well as the atmospheric boundary condition from \cite{Dziembowski1971}. The Eulerian perturbations to pressure, density, and gravitational potential for a particular eigenmode are a direct output from the \GYRE~code while they also can be obtained from the {displacement} eigenfunctions directly (\GYRE~and \ADIPLS).

\begin{figure*}
	\centering

	\includegraphics[width=0.92\linewidth]{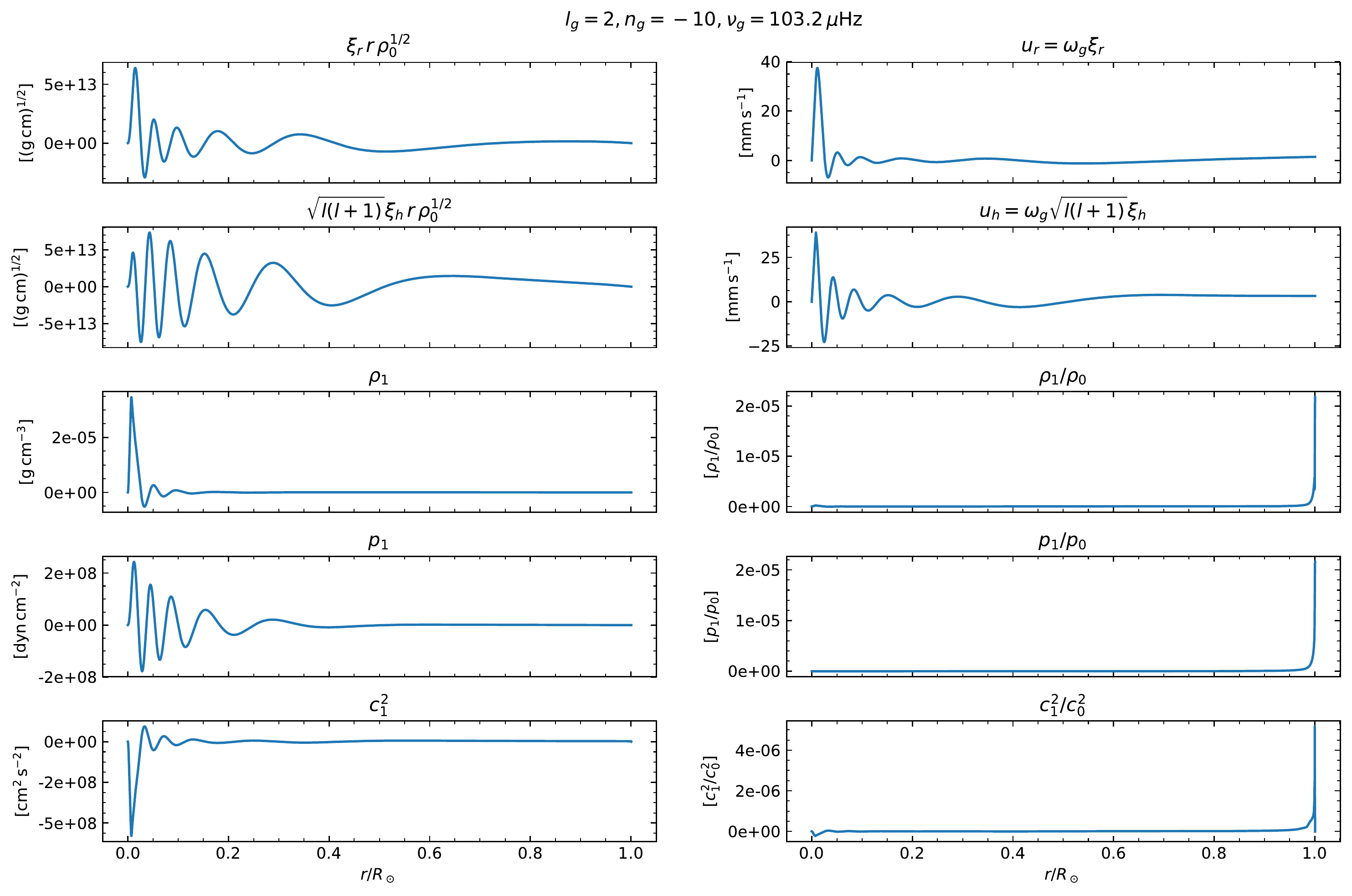}

	\caption{Eigenfunctions and {Eulerian} perturbations to solar structure for a typical g mode with $l_g=2,n_g=-10,\nu_g=103.2\,\mu\rm{Hz}$ \citep[compare to Fig.~5.10 in][]{JCDoscilAsBook}. The mode was normalised to a surface amplitude of $1\,\rm{mm}\,\rm{s}^{-1}$, which is at the order of observational upper limits  \citep[e.g.][]{Appourchaux2010}. Displayed are (from top to bottom) radial and horizontal eigenfunctions, and the Eulerian perturbations to density, pressure, and squared sound speed. The left column shows that the kinetic energy density and absolute changes to solar structure caused by the mode are centred near the solar core. The right column shows that relative changes to solar structure are very small, and in the case of density, pressure, and sound speed, reach the largest values near the surface. The exact behaviour of the relative Eulerian changes to solar structure at the surface depends on the choice of boundary condition used in the eigenfunction computation.}
	\label{figExampleEulg}
\end{figure*}

\begin{figure*}
	\centering

	\includegraphics[width=0.92\linewidth]{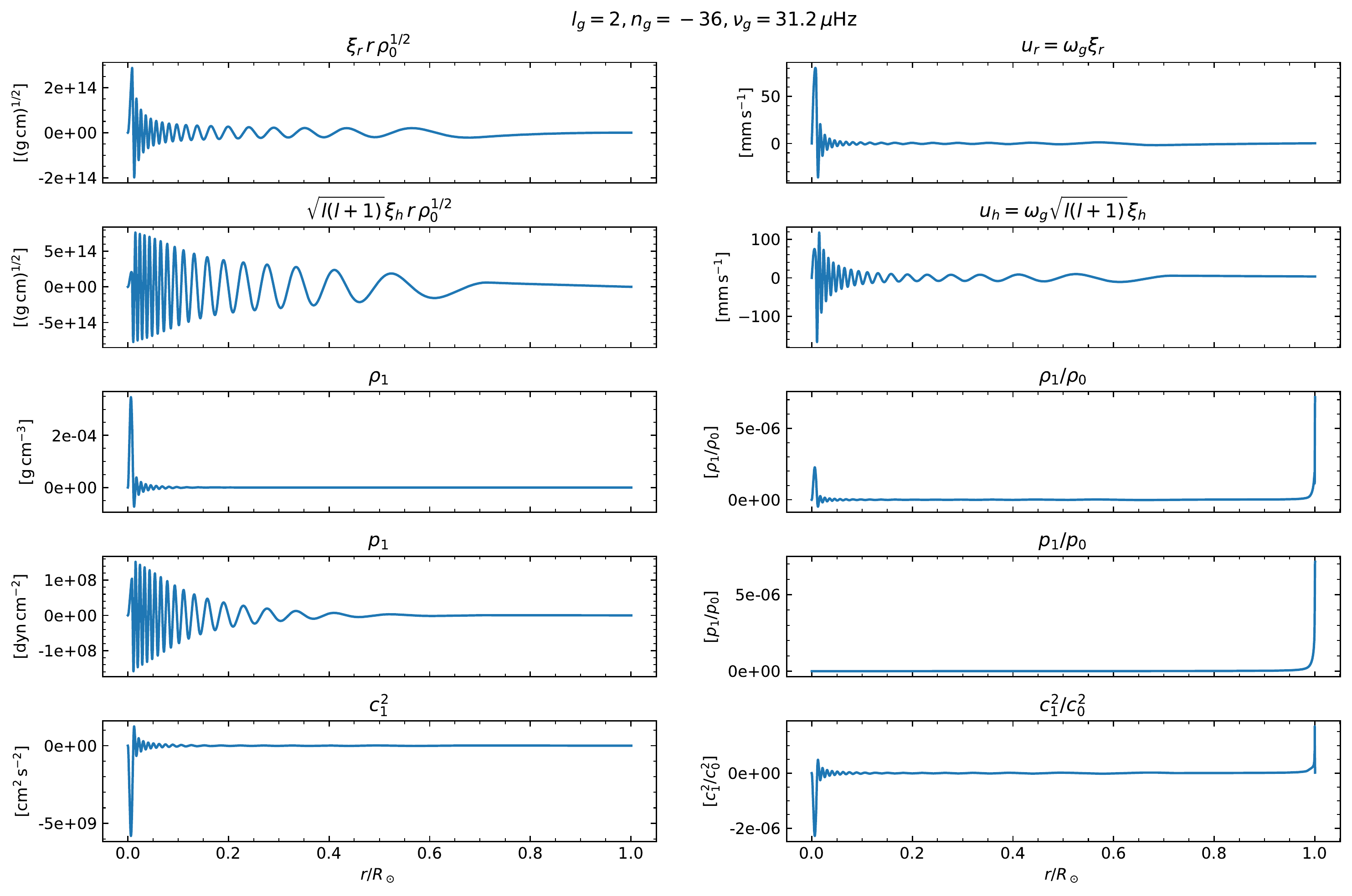}

	\caption{Same as Figure~\ref{figExampleEulg}, but for a g mode with $n_g=-36,\nu_g=31.2\,\mu\rm{Hz}$. See also Figures~\ref{figAsymptEf} and~\ref{figAsymptEfStruct}.}
	\label{figExampleEulg36}
\end{figure*}

As an example, we show eigenfunctions for $l_g=2$ g modes with $n_g=-10$ and $n_g=-36$ in Figures~\ref{figExampleEulg} and~\ref{figExampleEulg36}. The eigenfunctions were computed using the slightly smoothed version of model S, the \GYRE~code, and the boundary condition of an isothermal atmosphere from \cite{JCDadipls}. The panels in the left columns of Figures~\ref{figExampleEulg} and~\ref{figExampleEulg36} show that the kinetic energy density of the g modes is concentrated in the radiative zone. The g modes are trapped in this region, the eigenfunctions are oscillatory and the absolute changes to solar structure variables caused by the g mode, i.e. Eulerian perturbations to density, pressure, and squared sound speed $c^2$, reach their maximum amplitude. In the convection zone, the amplitude of the kinetic energy density is lower and decays. The mode is evanescent in this region. This is well known \citep[e.g.][]{JCDoscilAsBook}.

On the other hand, the panels in the right columns of Figures~\ref{figExampleEulg} and~\ref{figExampleEulg36} show that relative changes to density, pressure, and sound speed are very small, namely at the order of $10^{-6} - 10^{-5}$ for a g mode normalised to a surface amplitude of $1\,\rm{mm}\,\rm{s}^{-1}$. This is close to the upper limit for the g-mode amplitude from observations \citep[e.g.][and references therein]{Appourchaux2010}. Even more interestingly, the relative changes to solar structure caused by the g mode reach local maxima close to the solar surface and their amplitude is larger than in the core, at least for g modes of {lower} radial order. For modes with high radial order, the relative changes to solar structure are larger but of similar order of magnitude in the core compared to the surface. This concentration of relative changes to solar structure near the surface is due to the sharp decrease of density and pressure in this region and takes place despite the concentration of kinetic energy near the core {and despite the evanescent character of the g modes in the convection zone}. To guarantee the correctness of the eigenfunction computation, we have performed a number of tests, see Appendix~\ref{appEF}. These tests include a comparison of eigenfunctions to asymptotic formulae, see Figures~\ref{figAsymptEf} and~\ref{figAsymptEfStruct}.

\pdfsuppresswarningpagegroup=1

The question is now, how p modes are sensitive to the perturbation to solar structure from a g mode. 
It is well known that p modes are trapped between their lower turning point, which mostly depends on their horizontal surface phase speed $\omega R_\sun/\sqrt{l(l+1)}$, and their upper turning point close to the photosphere \citep[e.g.][]{JCDoscilAsBook}. The kinetic energy of p modes is concentrated in this region and it is increasing towards the surface. In addition, p modes spend most of their {life-}time close to the surface due to the strong increase of the sound speed in the solar interior.

As a consequence, the sensitivity of p modes to relative changes in solar structure increases towards the surface \citep[e.g. Figures~10 and~11 in][]{Basu2016}. At the same time, relative changes to solar structure due to g modes reach considerable amplitudes near the surface as we have illustrated in Figures~\ref{figExampleEulg} and~\ref{figExampleEulg36}.

We thus conclude that the coupling of a g mode with a p mode takes place to a large extent near the solar surface, if there is any such coupling present in the Sun. The hypothesis that the increase of p-mode sensitivity in the convection zone would overcompensate the decrease in amplitude of the g mode was already put forward by \citet{Wentzel1987}.

\section{P-mode frequency perturbation due to a frozen g mode}
\label{secperturbtheory}

In the following, we derive equations for the first-order perturbation to a p-mode eigenfrequency in the presence of a g mode. To do that, we assume that the period of the g mode is much longer than that of the p mode. Furthermore, we assume that the g-mode period is long compared to the observational time scale. To simplify the analysis, we therefore assume that the perturbation to solar structure caused by a g mode is constant in time \citep[see also][]{Kennedy1993,ScherrerGough2019}. In this picture, the time-dependence of the g mode may be added after the frequency shift has been calculated.

The perturbed wave operator for changes to solar structure has been derived before \citep[see, e.g.][and references therein]{Basu2016}. However, these derivations were made for a perturbation to solar structure that preserves hydrostatic equilibrium. As the perturbation in our case is due to a g-mode oscillation, this assumption is no longer valid. We therefore rederive the perturbed wave operator for a perturbation to solar structure that does not necessarily conserve hydrostatic equilibrium, see Appendix~\ref{appPerturbedWaveOp}.

The perturbed wave operator $\cL + \delta \cL$ acts on a p-mode with eigenfunction $\bxi_p$ and frequency $\omega_p$, where the index $p$ stands for a p mode with
{\begin{align}
p&=(l_p,m_p,n_p)=(l,m,n), \quad n_p>0.
\end{align}}
On the other hand, the perturbation is caused by a g mode with eigenfunction $\bxi_g$ and frequency $\omega_g$ with index
{\begin{align}
g=(l_g,m_g,n_g)=(l',m',n'), \quad n_g<0.
\end{align}}
{Consequently, we can express changes to all structure quantities, namely density, pressure, gravitational acceleration, and squared sound-speed, as well as the flow field associated to the g-mode oscillation in terms of the g-mode eigenfunction or Eulerian perturbations} (see also Equations~[\ref{eqp1}]-[\ref{eqg1}]),
\begin{align}
\delta p_0 &= p_{1,g}, \label{eqdp1}\\
\delta \rho_0 &= \rho_{1,g} , \\
\delta \bg_0 &= \bg_{1,g} , \\
\delta(c_0^2) &= -\frac{\Gamma_{1,0} \,p_0}{\rho_0^2} \delta\rho_0  + \frac{\Gamma_{1,0}}{\rho_0} \delta p_0  {+} \frac{p_0}{\rho_0} \delta \Gamma_{1,0}, \\
\delta \Gamma_{1,0} &= {-} \nabla \Gamma_{1,0} \cdot \bxi_g, \label{eqdGamma}\\
 \delta \bu_0 & =  -\ii\omega\bxi_g  \label{equ} ,
\end{align}
where $\Gamma_{1,0}(r)$ is the first adiabatic exponent of the equilibrium solar model \citep{JCDadipls,Aerts2010}. {In Equation~\eqref{eqdGamma}, we have assumed that the Lagrangian perturbation to the first adiabatic exponent during the g-mode oscillation is zero. The adiabatic exponent is a material property and reflects the chemical composition and the ionization state of the plasma \citep[see, e.g.][chapter 14]{KippenhahnWeigert1990Book}. During the oscillation, the chemical composition does not change when moving with the plasma. The degree of ionization, however, is a function of the gas pressure and temperature of the plasma. As temperature and pressure do change during the oscillation, the degree of ionization can change as well. This may have a small effect on the adiabatic exponent in the ionization zones of the Sun, which we neglect in this analysis.}

\subsection{Frequency shifts in the rotating Sun}

{The} p-mode eigenfunctions $\hat\bxi_p$ of the wave equation \eqref{eqwave} for the rotating Sun can be {expressed} in terms of the solutions of the eigenfunctions in the non-rotating case, $\bxi_p${, to first order in quasi-degenerate perturbation theory \citep[][]{Lavely1992}}, according to
\begin{align}
\hat\bxi_p &= \bxi_p + \hat \bxi_p^{(1)}, \label{eqpperturb}\\
\hat \bxi_p^{(1)} &=\sum_{\substack{p'\neq p \\ m_{p^{'}}=m_p}} c_{p,p'} \bxi_{p'}, \label{eqpperturbation} \\
c_{p,p'} &= \frac{-2\ii\omega_p\langle \bxi_{p^{'}}, (\bu_0 \cdot \nabla) \, \bxi_p \rangle}{\omega_p^2 - \omega^2_{p'}} , \label{eqpperturbcoeff}
\end{align}
where we note that in Equation~\eqref{eqpperturbation} only terms of the same $m$ as in $\bxi_p$ enter. The perturbation to the eigenfunction, $\hat \bxi_p^{(1)}$, has a much smaller amplitude than the unperturbed eigenfunction if the harmonic degree $l_p$ of the unperturbed mode is small \citep[compare to][]{Woodard1989}. Similarly, the perturbed eigenfrequencies can be written, to first order, as \citep[e.g.][]{Ritzwoller1991}
\begin{align}
\hat \omega_{p} = \omega_{p}  + \hat\omega_{p}^{(1)}. \label{eqperturbfreq}
\end{align}
We assume that rotation does not couple p modes and g modes since their frequencies differ largely.

For g modes, the same expansion as in Equations~\eqref{eqpperturb}-\eqref{eqperturbfreq} is possible, and we note that the perturbation to the eigenfunction, $\hat \bxi_g^{(1)}$, has a much smaller amplitude than the eigenfunction of the non-rotating model also for g modes in the case of small harmonic degrees as considered in this paper.

Given the eigenfunctions and mode frequencies for the rotating solar model, we can now evaluate the frequency perturbation of the p mode due to a g mode using non-degenerate perturbation theory. The use of non-degenerate perturbation theory will be justified later because the frequency shift from the g modes is much smaller than any frequency difference between the eigenfrequencies of the rotating eigenmodes. In this case, the frequency shift from a g mode is simply given by
\begin{align}
\delta (\hat \omega_p^2) &= \frac{\langle \hat\bxi_p, \delta \hat\cL [\hat\bxi_p] \rangle}{\langle \hat\bxi_p, \hat\bxi_p \rangle},\label{eqFreqshiftRotBackground}
\end{align}
where $\delta\hat{\cL}=\delta\cL + \delta\hat\cL^{(1)}$ indicates that the perturbation is caused by a g mode in a rotating Sun, $\hat\bxi_g=\bxi_g + \hat \bxi_g^{(1)}$. Using this and $\langle \hat\bxi_p, \hat\bxi_p \rangle=1+\langle \hat\bxi_p^{(1)}, \hat\bxi_p^{(1)}\rangle$ (see Equations~[\ref{eqeforthonormal}] and~[\ref{eqpperturb}]), Equation~\eqref{eqFreqshiftRotBackground} can be written to leading order as
\begin{align}
\delta (\hat \omega_p^2) 
&= \langle \bxi_p, {\delta \cL} [\bxi_p] \rangle 
= \delta (\omega_p^2). 
\label{eqFreqshiftRotBackground3} 
\end{align}
Here, all terms involving $\hat\bxi_p^{(1)}$ or $\delta\hat\cL^{(1)}$ are of higher than first order and have therefore been omitted. 
Hence, we evaluate $\delta(\omega_p^2)$ and the frequency shift as $2\pi\,\delta\nu_p=\delta\omega_p= \delta(\omega_p^2)/(2\omega_p)$ in the following.

The scalar product in Equation~\eqref{eqFreqshiftRotBackground3} leads to a number of integrals over one g-mode and two p-mode eigenfunctions, which are given in detail in Appendix~\ref{appPerturb} for the terms involved in $\delta\cL$. Similar integrals have been found by other authors for the coupling of g and p modes \citep[e.g.][]{Dziembowski1982,Kennedy1993,ScherrerGough2019}{, as well as for flows and} aspherical perturbations to solar structure preserving hydrostatic equilibrium {that are decomposed in spherical harmonics \citep[e.g.][]{Lavely1992,Gough1993}}. We note here that we omitted one term ($\delta\cL_4$) {for simplicity}. This term models the change of the Eulerian perturbation to the gravitational potential during the p mode oscillation, or in other words changes to $\Phi_{1,p}$, in the presence of the g mode. This term is expected to have a small contribution to the wave equation.

\subsection{Selection rules}
\label{secSelectionRules}

Our derivation of the first-order frequency shift gives rise to selection rules for mode coupling, which are derived in Appendix~\ref{appHorizInt}. The horizontal integrals over three spherical harmonics appearing in Appendix~\ref{appPerturb} can be evaluated using Gaunt's formula (e.g, \citealp{Dahlen1998GlobalSeismology}, see also \citealp{Hanasoge2017FlowSoundSp,Kiefer2017b,Fournier2018}){, see Appendices~\ref{appHorizInt} and~\ref{appGaunt}}.

We find that a first-order frequency shift of a p mode by a g mode is possible only if all of the following conditions are satisfied.
\begin{enumerate}[I.)]
	\item The harmonic degree $l_g$ of the g mode is even.
	\item The azimuthal order $m_g$ of the g mode is zero.
	\item The harmonic degree $l_g$ of the g mode is not larger than twice the harmonic degree of the p mode ($l_g\leq 2 l_p$).
\end{enumerate}

\citet{Kennedy1993} and \citet{ScherrerGough2019} have found previously that only g modes of even harmonic degree produce a p-mode frequency shift at first order, see also \citet{Gough1993} for perturbations to solar structure that preserve hydrostatic equilibrium. 
Although \citet{ScherrerGough2019} do not note a selection rule regarding $m_g$, it can be derived using their framework of degenerate perturbation theory{, where both the g mode and rotation where treated as perturbations to the solar model}.

This can be seen by considering the coupling matrix that leads to the perturbed p-mode eigenfunctions and perturbed p-mode frequencies at first order. If $m_g\neq 0$, the diagonal of that matrix is equal to the frequency perturbation from rotation. The coupling due to g modes appears at the off-diagonal matrix elements where $m_p - m_{p'} = m_g$. In the notation of \citet{ScherrerGough2019}, this corresponds to $\mu-\mu'=m$, see Equations (25) - (27) in their work. All other matrix elements are equal to zero. As the off-diagonal matrix elements due to the g mode are much smaller than the diagonal matrix elements due to rotation, the off-diagonal matrix elements only lead to a second order perturbation in the eigenfrequency and the first-order frequency shift from the g mode is zero.

Somewhat similarly, the selection rules I-II can be obtained using the framework of quasi-degenerate perturbation theory outlined in \citet[][see also~\citealp{SchadPhD}]{Lavely1992}.

Here, we have modelled the first-order frequency shift in the p modes but we did not model the change in p-mode travel time as measured by \cite{Fossat2017}. Given the above selection rules, global p-mode frequency shifts can be ruled out as a cause for the observations of \cite{Fossat2017}. However, in our eyes, it is not straightforward to make a direct connection between frequency shifts and the measurement of the round-trip travel time by \citet{Fossat2017}. For example, it is in principle possible that perturbations to the p-mode eigenfunctions cause a signature of g modes in this measurement.

\section{Numerical results}
\label{secResults}

In the following, we numerically evaluate the analytical expression for the first-order perturbation to the p mode eigenfrequency that we derived in the preceding section. Here, we present results for g modes with harmonic degree $l_g=2$ and p modes with $l_p=2$. {Results for $l_p=1$ are very similar.} The g mode eigenfunctions are normalised to a surface amplitude of $1\,\rm{mm}\,\rm{s}^{-1}$, which is at the order of observational upper limits for the amplitude of solar g modes \citep[e.g.][]{Appourchaux2010}. The frequency shifts computed in this section can therefore either be seen as approximative upper limits, or they can be rescaled to different g-mode surface amplitudes.

Inspired by the analysis of \cite{Fossat2017}, we assume that the observational time scale is eight hours and therefore apply our results to g modes with periods longer than that, which means frequencies $\nu_g <34.7\,\mu\rm{Hz}$. Moreover, we obtain results also for low-order g modes with frequencies above $34.7\,\mu\rm{Hz}$ for information purposes, even though our assumptions of a clear separation of observational time scale and g-mode period may break down in this range. We do not consider results obtained for $n_g<-200$ ($\nu_g < 5.67\,\mu\rm{Hz}$) in the following. We checked however, that this limitation does not effect the major conclusions presented here. In addition to the p-mode range considered by \cite{Fossat2017}, we obtain results for p modes with frequencies $\nu_p < 5\,\rm{mHz}$. The results presented in this section were obtained from eigenmodes computed with \GYRE~and solar model S \citep{JCD1996}, and we assumed an isothermal atmosphere as surface boundary condition \citep{JCDadipls}, see also Appendix~\ref{appBC}.

\subsection{The effect is tiny in magnitude}

\begin{figure*}
	\centering

\includegraphics[width=\linewidth]{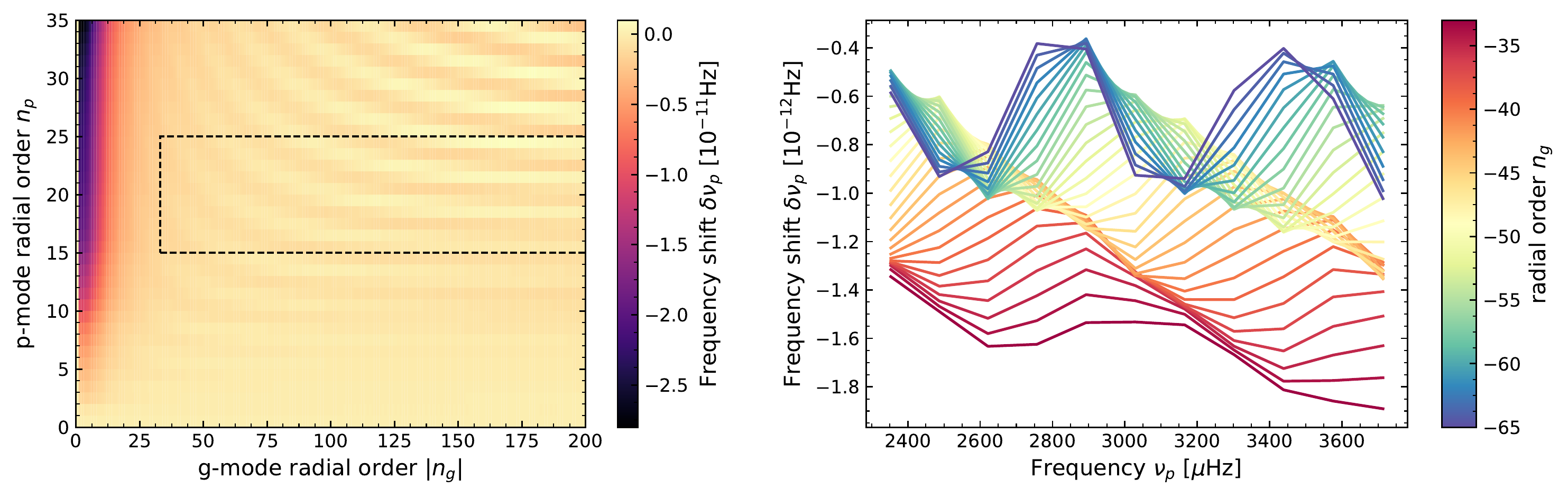}

	\caption{{First-order perturbations to the p-mode frequency due to a g-mode oscillation assumed to be frozen over the observation period. Results are shown for g modes and p modes with harmonic degrees $l_g=l_p=2$ and azimuthal orders $m_g=m_p=0$. {\it Left:} Frequency shift as a function of radial order for mode frequencies $\nu_p<5\,\rm{mHz}$ and $\nu_g>5.67\,\mu\rm{Hz}$. The range of modes studied by \citet{Fossat2017}, \citet{Fossat2018}, and \citet{Schunker2018} is inside the dashed rectangle, extending outwards of the plotted region at $|n_g|>200$. {\it Right:} Cuts through the results shown in the dashed region in the left panel for a selection of g-mode radial orders.} 
	}
	\label{figfreqshift}
\end{figure*}

As a first result, we find that the magnitude of the frequency shift of a p mode caused by a frozen g mode is extremely small, see Figure~\ref{figfreqshift}. All frequency shifts $\delta\nu_p$ computed are found to be below $0.1\,\rm{nHz}$. The relative frequency shifts $\delta\nu_p/\nu_p$ are all below $2\times 10^{-8}$. These numbers are in very broad agreement with the estimates presented in \citet{ScherrerGough2019}, who reported an order of magnitude of $2 \times 10^{-8}$ when converted to the g-mode amplitude assumed in this work.

Generally, we find the frequency shifts to be greatest for large p-mode and large g-mode frequencies. The study by \cite{Fossat2017} considered a range of g mode frequencies, which does not cover the g modes with the largest frequencies, see Figure~\ref{figfreqshift} (right panel and dashed region in left panel). For these modes, the p mode frequency shifts are smaller by about one order of magnitude and do not exceed $0.01\,\rm{nHz}$, or relative shifts of ${10^{-9}}$.

The observational upper limit on the g-mode amplitude depends on mode frequency, see for instance Figure~15 in \citet{Appourchaux2010}. While it is around $1\,\rm{mm}\,\rm{s}^{-1}$ for the g modes with the lowest radial orders, it is at the order of $1\,\rm{cm}\,\rm{s}^{-1}$ for modes with moderately higher radial orders, for example $\nu_g\approx 35\,\mu\rm{Hz}$, $n_g\approx - 33$. This is where the analysis of \citet{Fossat2017} is sensitive to. Even if we rescale the frequency shifts for all modes with $\nu_g <  35\,\mu\rm{Hz}$ to an amplitude of $1\,\rm{cm}\,\rm{s}^{-1}$, the maximum relative frequency shift among all modes does not change. This can also be seen by inspecting the relevant region in Figure~\ref{figfreqshift}.

\subsection{Contributions from density, pressure, sound speed, and flow}

\begin{figure}
	\centering

	\includegraphics[width=1\linewidth]{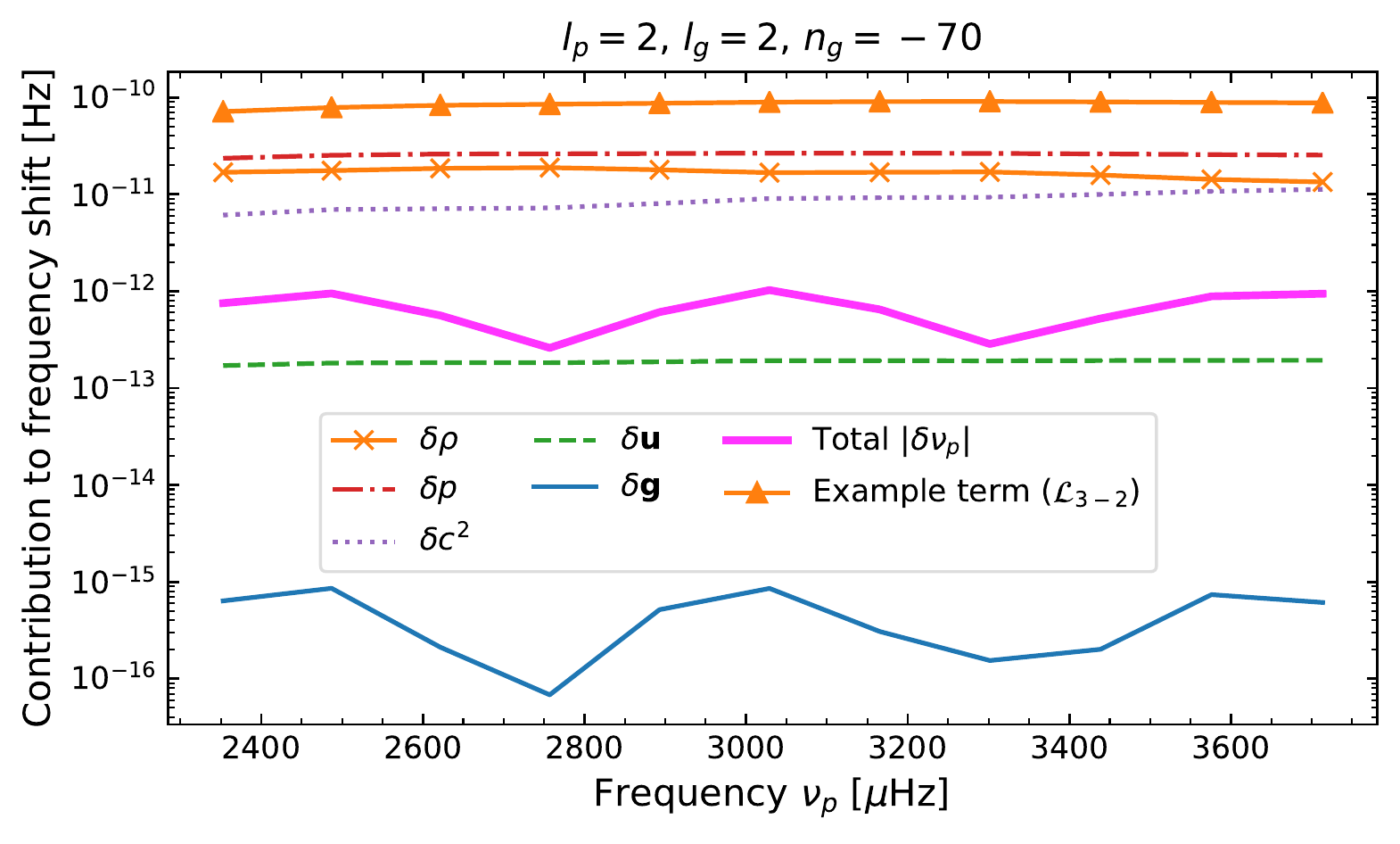}

	\caption{Contribution to the frequency shift of p modes by changes to solar structure variables $\rho,p,c^2,\bg$, and the flow field $\bu$ caused by a g mode {in absolute values}. The total frequency shift {$|\delta \nu_p|$} and the contribution from an example term ($\delta\cL_{3-2}$, see Equation~[\ref{eq:int3-2}]) are also given for comparison. The azimuthal order of the modes is $m_p=m_g=0$. }
	\label{figfreqshiftcontrib}
\end{figure}

The question naturally arises, which of the effects of the g mode, the induced changes to density, pressure, sound speed, gravitional acceleration, or its flow field, is the dominant contributor to the frequency shift.
In order to answer this question, we sum the contributions from the individual terms derived in Appendix~\ref{appPerturb} for each of these quantities. {The results are shown in Figure~\ref{figfreqshiftcontrib}.} In addition, we also show the contribution of {an example term that usually is of large amplitude ($\delta\cL_{3-2}$, see Eq.~[\ref{eq:int3-2}]), and which is due to changes in density.} It can be seen that this term is usually about one order of magnitude larger than the total density contribution.

Similarly, the total contribution from relevant structural changes, that is from density, pressure, and sound-speed, is again about an order of magnitude smaller than their individual contribution. This is because these terms tend to cancel each other to a large extent. The contribution from changes to gravity is negligible.

Even though the contribution of the flow is about two orders of magnitude smaller than from individual density terms, it still has a relevant contribution to the total frequency shift of order 10~\%. In some cases, this number is larger, if the total frequency shift is close to zero.

In general, however, the effect of structural changes dominates over the effect of a flow caused by the g mode.

\subsection{Regions of interaction}

\label{secSurface}

\begin{figure*}
	\centering

	\includegraphics[width=1\linewidth]{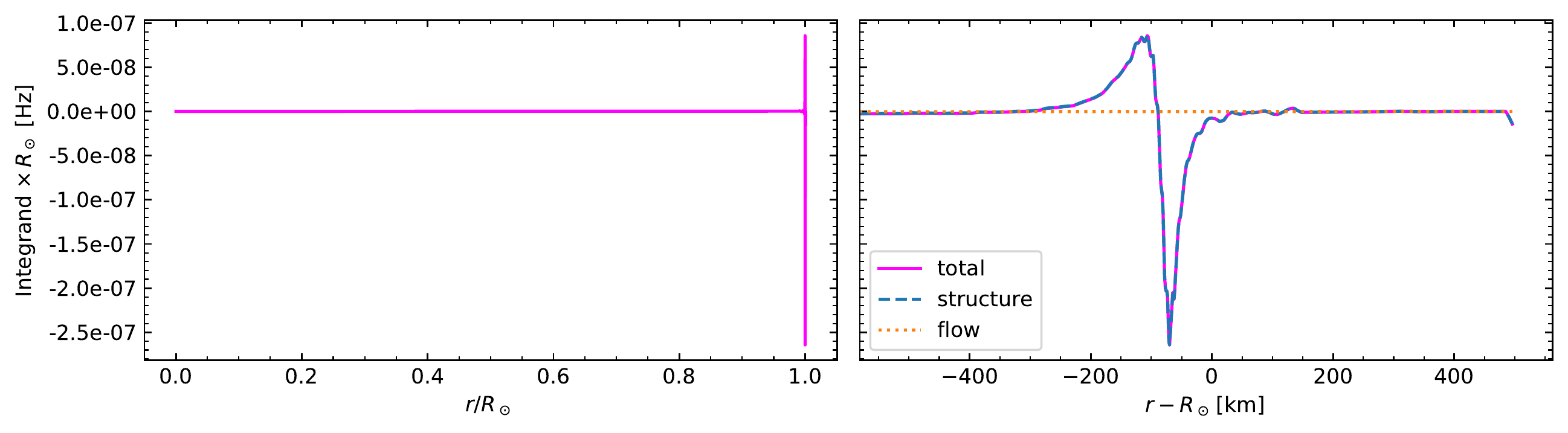}

	\caption{Radial integrands for computing the frequency shift of a p mode with $l_p=2,m_p=0,n_p=20$ caused by a g mode with $l_g=2,m_g=0,n_g=-36$. When integrating over radius, the total integrand yields the total frequency shift $\delta \nu_p$ (both panels). The contributions to the total integrand resulting from the structural changes and the flow are also shown (right panel). The spike at the surface boundary is due to numerical difficulties in computing second order derivatives at the boundary. Therefore, the three outermost radial grid points were discarded.}
	\label{figIntegrand}
\end{figure*}

We further investigate in which region in the Sun the interaction predominantly occurs. To do that, we rewrite the total integral leading to the frequency shift from Equation~\eqref{eqFreqshiftRotBackground3} as one radial integral. In addition, we decompose this integrand in the contribution from changes to solar structure and from the flow, see Figure~\ref{figIntegrand}. 
We find that the total integrand takes the largest values very close {to} the surface and that it is nearly identical to the integrand from the structure terms. This is in agreement with our finding from Section~\ref{secgmodeef} that the relative changes to solar structure from a g mode are concentrated near the surface. However, the integrands change sign near the surface, and the total contribution from the near-surface layers is thus smaller than the magnitude of the integrands suggest.

\begin{figure}
	\centering

	\includegraphics[width=1\linewidth]{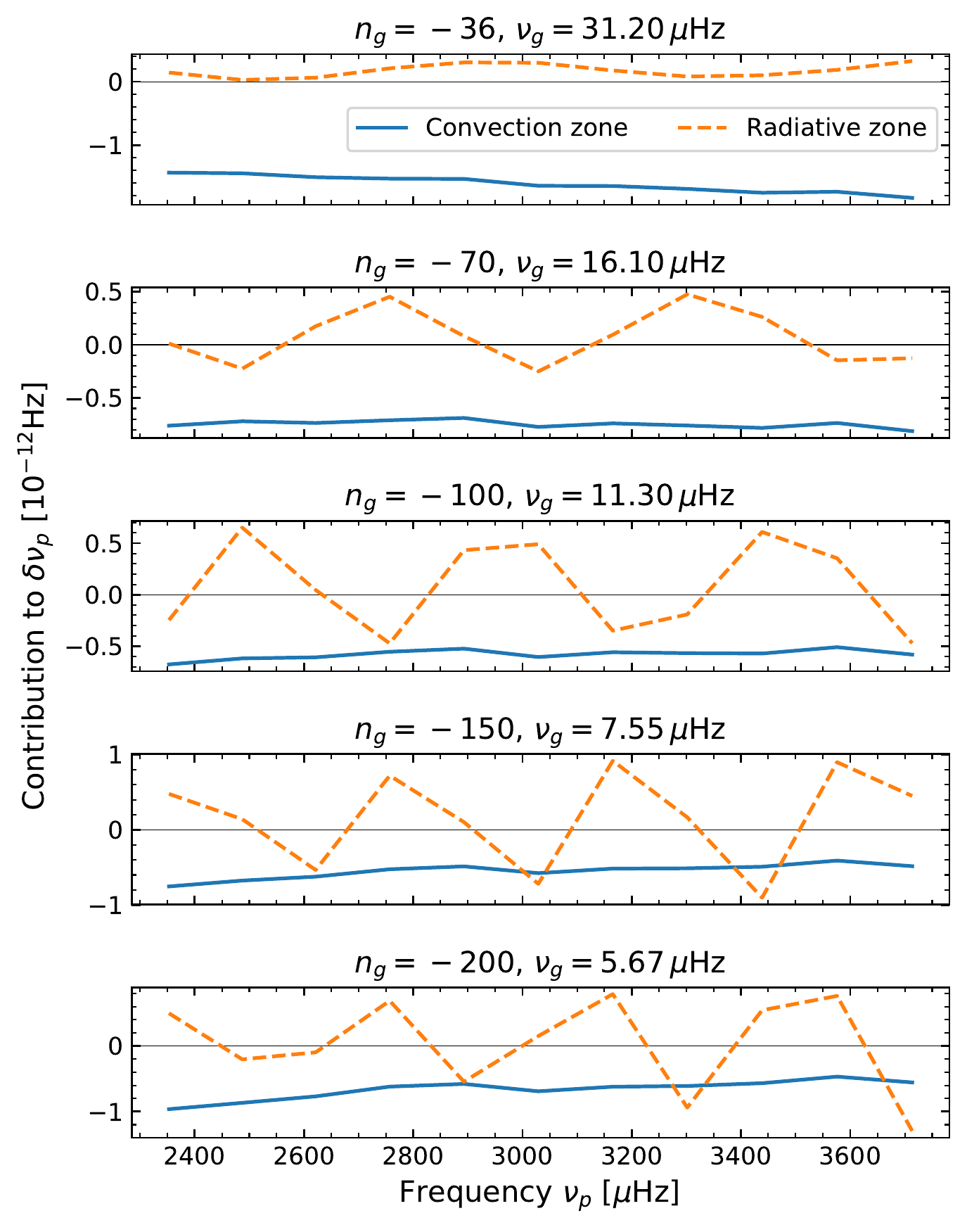}

	\caption{Contribution to the frequency shift from the convection zone and from the radiative zone as a function of p-mode frequency for several typical g modes considered in this study. Results are shown for $l_p=l_g=2$, $m_p=m_g=0$.}
	\label{figContribCZ}
\end{figure}

To investigate this further, we computed the contributions to the total integral from the convection zone and from the radiative zone, which are shown in Figure~\ref{figContribCZ}. It can be seen that the contribution from the radiative zone is about an order of magnitude smaller than from the convection zone ($n_g=-36$ in Fig.~\ref{figContribCZ}). A similar conclusion was reached by \citet{Wentzel1987} by the simple argument that the evanescent behaviour of the g modes in the convection zone is overcompensated by the increase in sensitivity of the p modes. However, the contribution from the radiative zone can reach a comparable magnitude as from the convection zone for higher-order modes, for example $n_g\leq-100$, see Figure~\ref{figContribCZ}.

{Following \citet{JCD1997}, one may ask whether the above conclusions stay the same if we use Lagrangian instead of Eulerian perturbations to solar structure. These authors found that contributions to the frequency shift from, e.g, Eulerian changes to sound speed and density nearly cancel as it is the case in our study. If one uses Lagrangian perturbations, such a near cancellation does not occur as strongly in their computation.

We therefore recomputed our main results using Lagrangian perturbations to solar structure for a selected number of mode pairs as test cases. As expected, the resulting total frequency shifts agree with the ones from the Eulerian perturbations (successful for low radial order g-modes $|n_g|\lesssim 40$).

We further considered the example of quadrupolar modes with $n_g = - 36$ and $n_p = 20$ in more detail. For this pair, the radial integrand from the Lagrangian variables has the same order of magnitude as the Eulerian one and a qualitatively similar behaviour. Both assume maximal values in the same region close to the surface. Their actual functional dependence on depth is different, which is consistent with \citet{JCD1997}. The contributions from the radiative zone and the convection zone are of comparable amplitude in the Lagrangian case, but of opposite sign. These contributions have a larger amplitude as in the Eulerian case, which leads to additional near-cancellation. This is found for all pairs of modes we considered. Thus, Lagrangian perturbations do not seem to us as a better choice for studying this problem.

For both Eulerian and Lagrangian variables, we find in general that the contribution from near-surface layers is of similar magnitude as the contribution from the convection zone. We thus conclude that the interaction of g and p modes is taking place to a large part near the surface, whether solar structure perturbations are computed in Eulerian or Lagrangian variables.
}

{
\subsection{Dependence on solar model and surface boundary conditions}
\label{secDependenceOnModelandBC}

Due to the large magnitude of the integrands near the surface, one may of course ask} whether and how the numerical results depend on the surface boundary condition applied in the eigenfunction computation.

We therefore computed eigenfunctions using {a number of different boundary conditions, which are implemented in the \GYRE~code and} which are detailed in Appendix~\ref{appBC}.

In addition, we checked whether our results depend on the solar model used to compute the eigenfunctions. The second derivative of the equilibrium density enters in our results, although it is not quite smooth in the original version of model S. This leads to spikes in some of the integrands near the core and wiggles near the surface, see for example the right panel in Figure~\ref{figIntegrand}. We therefore also computed eigenfunctions using a version of model S, where the density was slightly smoothed, see details in Appendix~\ref{appBC}. In order to get a rough idea about the magnitude of possible systematic errors, we also obtained results for a $1\,M_\sun$ stellar model obtained using \MESA~\citep{Paxton2011}.

We find that the numerical results depend to some extent on the boundary condition and solar model used. 
Qualitative results, however, are unchanged, like the overall magnitude of the effect, selection rules, the interaction taking place to a large part near the surface, that the effects of density, pressure and sound speed nearly cancel, and that the changes to solar structure have a larger effect than the flow.

For all boundary conditions and models, the general pattern and the order of magnitude is the same as in Figure~\ref{figfreqshift}. For some boundary conditions, models and mode combinations, however, parts of the pattern visible in the right panel in Figure~\ref{figfreqshift} are shifted by about $0.5 - 1.0 \times 10^{-12}\,\rm{Hz}$.

There are a number of good reasons why the observed sensitivity to solar model and surface boundary conditions is not surprising. Firstly, we found that the interaction of p and g modes takes place {to a large part} near the surface. This layer has thus to be accurately modelled to guarantee an accurate answer. Secondly, {p modes are sensitive to} relative perturbations to solar structure. Close to the surface, the relative Eulerian perturbations from a g mode depend on the surface boundary conditions used in the eigenfunction computation. Thirdly, we found that the effects of pressure, density, and sound speed almost cancel, with the final frequency shift being nearly two orders of magnitude smaller than the largest term involved in the sum. Substracting terms of similar magnitude may result in an amplification of systematic errors.

As a consequence, all effects have to be modelled in a very accurate way if quantitative predictions from this method are to be used.

Finally, it is unknown whether convection has effects on g modes that are not included in the eigenfunction computation such as by turbulence. For example, if the amplitude of the g-mode eigenfunction was significantly reduced at the surface compared to the core, the contribution of the radiative zone to the frequency shifts would be enhanced. Attenuation of p and g modes has not been included in our model.

\section{Conclusion}
\label{secConclusion}

In this paper, we have developed a model for the first-order frequency shift of solar p modes caused by solar g modes in the framework of non-degenerate time-independent perturbation theory. To do so, the time-dependent perturbation to solar structure caused by a g mode is assumed to be constant at the observational time scale and the degeneracy in mode frequency is assumed to be lifted by rotation.

Our study was motivated by a recent study by \cite{Fossat2017}, who proposed to detect g modes by measuring  their impact on the wave travel time across the Sun, which is inversely proportional to the large frequency separation of p-mode frequencies.

Firstly, we find that indeed g modes can perturb global p-mode eigenfrequencies. However, the effect is extremely small. For a g-mode surface amplitude of $A_g=1\,\rm{mm}\,\rm{s}^{-1}$, all frequency shifts computed with our model are smaller than 
$0.1\,\rm{nHz}$, or in relative numbers
\begin{align}
    \frac{|\delta \nu_p|}{\nu_p} \leq 2\times 10^{-8} \frac{A_g}{1\,\rm{mm}\,\rm{s}^{-1}},
\end{align}
where {results for p modes with $l_p=1,2$ and g modes with $l_g=2$ are taken into account}. Measuring such {a small effect would require an extremely long observation time}. Our results are in broad agreement with the estimates obtained by \citet{ScherrerGough2019}.

Secondly, we find that the interaction of g and p modes takes place to a large part in near-surface layers. This is due to two reasons. On the one hand, p modes are most sensitive to relative changes to solar structure close to the surface. On the other hand, relative changes caused by g modes are relatively large in this region, too. This is the case despite the large amplitude of g modes in the solar core, and caused by the sharp decrease of pressure and density near the surface. Furthermore, the effect from structural changes dominates over the effect from the flow caused by the g mode.

Thirdly, we find that a first-order frequency shift of a p mode by a g mode is only possible if all of the following selection rules are satisfied. The harmonic degree $l_g$ of the g mode has to be even, the azimuthal order $m_g$ of the g mode has to be zero, and the harmonic degree $l_g$ of the g mode has to be smaller than or equal to twice the harmonic degree of the p mode ($l_g\leq 2 l_p$). The condition that the harmonic degree of the g mode has to be even has been previously obtained by \cite{Kennedy1993} and \citet{ScherrerGough2019}.

We caution that we only modelled the first-order frequency shift in the p modes but not the change in p-mode travel time as {measured} by \cite{Fossat2017}. {For example, we did not model the effect of g-modes on the p-mode eigenfunctions, which may result in different selection rules.}

While challenging, the exploration of various methods for detecting g modes certainly {remains} a worthwhile endeavour.

\begin{acknowledgements}
We thank Jesper Schou, Aaron Birch, and Damien Fournier for scientific discussions, Earl Bellinger and Warrick Ball for help with installing and running \GYRE, and Richard Townsend for answering questions on eigenfunction computations via the \GYRE~forum. {We thank the anonymous referee for a number of useful suggestions that improved the manuscript.} We used the tomso python package written by Warrick Ball for reading \ADIPLS~eigenfunctions and solar models. This work was supported in part by the Max Planck Society through a grant on PLATO Science. The computational resources were provided by the German Data Center for SDO through a grant from the German Aerospace Center (DLR). We acknowledge partial support from the European Research Council Synergy Grant WHOLE SUN \#810218. HH acknowledges the support of the Erasmus Mundus Joint Masters Programme, AstroMundus. Parts of this work were submitted as a MSc. thesis by HH.
\end{acknowledgements}

%-------------------------------------------------------------------
% Bibliography
%-------------------------------------------------------------------


\bibliographystyle{aa} % style aa.bst
\bibliography{myCompleteJabrefBiblio,library}

\begin{thebibliography}{54}
\expandafter\ifx\csname natexlab\endcsname\relax\def\natexlab#1{#1}\fi

\bibitem[{{Aerts} {et~al.}(2010){Aerts}, {Christensen-Dalsgaard}, \&
  {Kurtz}}]{Aerts2010}
{Aerts}, C., {Christensen-Dalsgaard}, J., \& {Kurtz}, D.~W. 2010,
  {Asteroseismology} (Dordrecht: Springer)

\bibitem[{{Appourchaux} {et~al.}(2010){Appourchaux}, {Belkacem}, {Broomhall},
  {Chaplin}, {Gough}, {Houdek}, {Provost}, {Baudin}, {Boumier}, {Elsworth},
  {Garc{\'{\i}}a}, {Andersen}, {Finsterle}, {Fr{\"o}hlich}, {Gabriel}, {Grec},
  {Jim{\'e}nez}, {Kosovichev}, {Sekii}, {Toutain}, \&
  {Turck-Chi{\`e}ze}}]{Appourchaux2010}
{Appourchaux}, T., {Belkacem}, K., {Broomhall}, A.-M., {et~al.} 2010, \aapr,
  18, 197

\bibitem[{{Appourchaux} \& {Pall{\'e}}(2013)}]{Appourchaux2013}
{Appourchaux}, T. \& {Pall{\'e}}, P.~L. 2013, in Astronomical Society of the
  Pacific Conference Series, Vol. 478, Fifty Years of Seismology of the Sun and
  Stars, ed. K.~{Jain}, S.~C. {Tripathy}, F.~{Hill}, J.~W. {Leibacher}, \&
  A.~A. {Pevtsov}, 125

\bibitem[{{Barrera} {et~al.}(1985){Barrera}, {Estevez}, \&
  {Giraldo}}]{Barrera1985}
{Barrera}, R.~G., {Estevez}, G.~A., \& {Giraldo}, J. 1985, European Journal of
  Physics, 6, 287

\bibitem[{Basu(2016)}]{Basu2016}
Basu, S. 2016, \lrsp, 13, 2

\bibitem[{{Berthomieu} \& {Provost}(1990)}]{Berthomieu1990}
{Berthomieu}, G. \& {Provost}, J. 1990, \aap, 227, 563

\bibitem[{{Berthomieu} \& {Provost}(1991)}]{Berthomieu1991}
{Berthomieu}, G. \& {Provost}, J. 1991, \solphys, 133, 127

\bibitem[{{Chris\-ten\-sen\mbox{-}Dals\-gaard}
  {et~al.}(1996){Chris\-ten\-sen\mbox{-}Dals\-gaard}, {Dappen}, {Ajukov},
  {Anderson}, {Antia}, {Basu}, {Baturin}, {Berthomieu}, {Chaboyer}, {Chitre},
  {Cox}, {Demarque}, {Donatowicz}, {Dziembowski}, {Gabriel}, {Gough},
  {Guenther}, {Guzik}, {Harvey}, {Hill}, {Houdek}, {Iglesias}, {Kosovichev},
  {Leibacher}, {Morel}, {Proffitt}, {Provost}, {Reiter}, {Rhodes}, {Rogers},
  {Roxburgh}, {Thompson}, \& {Ulrich}}]{JCD1996}
{Chris\-ten\-sen\mbox{-}Dals\-gaard}, J., {Dappen}, W., {Ajukov}, S.~V.,
  {et~al.} 1996, Science, 272, 1286

\bibitem[{Christensen-Dalsgaard(2003)}]{JCDoscilAsBook}
Christensen-Dalsgaard, J. 2003, Lecture Notes on Stellar Oscillations, 5th
  edition, available online at \url{http://users-phys.au.dk/jcd/oscilnotes/} on
  August 2, 2017.

\bibitem[{{Christensen-Dalsgaard}(2008)}]{JCDadipls}
{Christensen-Dalsgaard}, J. 2008, \apss, 316, 113

\bibitem[{{Christensen-Dalsgaard} \& {Thompson}(1997)}]{JCD1997}
{Christensen-Dalsgaard}, J. \& {Thompson}, M.~J. 1997, \mnras, 284, 527

\bibitem[{Dahlen \& Tromp(1998)}]{Dahlen1998GlobalSeismology}
Dahlen, F. \& Tromp, J. 1998, Theoretical Global Seismology (Princeton
  University Press)

\bibitem[{{Duvall}(2004)}]{Duvall2004}
{Duvall}, Jr., T.~L. 2004, in ESA Special Publication, Vol. 559, SOHO 14 Helio-
  and Asteroseismology: Towards a Golden Future, ed. D.~{Danesy}, 412

\bibitem[{{Dziembowski}(1971)}]{Dziembowski1971}
{Dziembowski}, W.~A. 1971, \actaa, 21, 289

\bibitem[{{Dziembowski}(1982)}]{Dziembowski1982}
{Dziembowski}, W.~A. 1982, \actaa, 32, 147

\bibitem[{{Fossat} {et~al.}(2017){Fossat}, {Boumier}, {Corbard}, {Provost},
  {Salabert}, {Schmider}, {Gabriel}, {Grec}, {Renaud}, {Robillot},
  {Roca-Cort{\'e}s}, {Turck-Chi{\`e}ze}, {Ulrich}, \& {Lazrek}}]{Fossat2017}
{Fossat}, E., {Boumier}, P., {Corbard}, T., {et~al.} 2017, \aap, 604, A40

\bibitem[{{Fossat} \& {Schmider}(2018)}]{Fossat2018}
{Fossat}, E. \& {Schmider}, F.~X. 2018, \aap, 612, L1

\bibitem[{{Fournier} {et~al.}(2018){Fournier}, {Hanson}, {Gizon}, \&
  {Barucq}}]{Fournier2018}
{Fournier}, D., {Hanson}, C.~S., {Gizon}, L., \& {Barucq}, H. 2018, \aap, 616,
  A156

\bibitem[{{Gabriel} {et~al.}(1995){Gabriel}, {Grec}, {Charra}, {Robillot},
  {Roca Cort{\'e}s}, {Turck-Chi{\`e}ze}, {Bocchia}, {Boumier}, {Cantin},
  {Cesp{\'e}des}, {Cougrand}, {Cr{\'e}tolle}, {Dam{\'e}}, {Decaudin},
  {Delache}, {Denis}, {Duc}, {Dzitko}, {Fossat}, {Fourmond}, {Garc{\'{\i}}a},
  {Gough}, {Grivel}, {Herreros}, {Lagard{\`e}re}, {Moalic}, {Pall{\'e}},
  {P{\'e}trou}, {Sanchez}, {Ulrich}, \& {van der Raay}}]{Gabriel1995}
{Gabriel}, A.~H., {Grec}, G., {Charra}, J., {et~al.} 1995, \solphys, 162, 61

\bibitem[{{Gough}(1993)}]{Gough1993}
{Gough}, D.~O. 1993, in Astrophysical Fluid Dynamics - Les Houches 1987, ed.
  J.-P. {Zahn} \& J.~{Zinn-Justin}, 399--560

\bibitem[{{Hamilton} \& {Blackstock}(1998)}]{Hamilton1998}
{Hamilton}, M.~F. \& {Blackstock}, D.~T., eds. 1998, Nonlinear acoustics
  (Academic press, San Diego)

\bibitem[{{Hanasoge} \& {Mandal}(2019)}]{Hanasoge2019}
{Hanasoge}, S. \& {Mandal}, K. 2019, \apjl, 871, L32

\bibitem[{{Hanasoge} {et~al.}(2017){Hanasoge}, {Woodard}, {Antia}, {Gizon}, \&
  {Sreenivasan}}]{Hanasoge2017FlowSoundSp}
{Hanasoge}, S.~M., {Woodard}, M., {Antia}, H.~M., {Gizon}, L., \&
  {Sreenivasan}, K.~R. 2017, \mnras, 470, 1404

\bibitem[{{Howe}(2009)}]{Howe2009}
{Howe}, R. 2009, \lrsp, 6, 1

\bibitem[{{Kennedy} {et~al.}(1993){Kennedy}, {Jefferies}, \&
  {Hill}}]{Kennedy1993}
{Kennedy}, J.~R., {Jefferies}, S.~M., \& {Hill}, F. 1993, in Astronomical
  Society of the Pacific Conference Series, Vol.~42, GONG 1992. Seismic
  Investigation of the Sun and Stars, ed. T.~M. {Brown}, 273

\bibitem[{{Kiefer} {et~al.}(2017){Kiefer}, {Schad}, \& {Roth}}]{Kiefer2017b}
{Kiefer}, R., {Schad}, A., \& {Roth}, M. 2017, \apj, 846, 162

\bibitem[{{Kippenhahn} \& {Weigert}(1990)}]{KippenhahnWeigert1990Book}
{Kippenhahn}, R. \& {Weigert}, A. 1990, {Stellar Structure and Evolution}
  (Springer-Verlag Berlin Heidelberg)

\bibitem[{{Kumar} \& {Goldreich}(1989)}]{Kumar1989}
{Kumar}, P. \& {Goldreich}, P. 1989, \apj, 342, 558

\bibitem[{{Lavely} \& {Ritzwoller}(1992)}]{Lavely1992}
{Lavely}, E.~M. \& {Ritzwoller}, M.~H. 1992, Philosophical Transactions of the
  Royal Society of London Series A, 339, 431

\bibitem[{{Liang} {et~al.}(2019){Liang}, {Gizon}, {Birch}, \&
  {Duvall}}]{Liang2019Rossby}
{Liang}, Z.-C., {Gizon}, L., {Birch}, A.~C., \& {Duvall}, T.~L. 2019, \aap,
  626, A3

\bibitem[{{L{\"o}ptien} {et~al.}(2018){L{\"o}ptien}, {Gizon}, {Birch}, {Schou},
  {Proxauf}, {Duvall}, {Bogart}, \& {Christensen}}]{Loeptien2018}
{L{\"o}ptien}, B., {Gizon}, L., {Birch}, A.~C., {et~al.} 2018, Nature
  Astronomy, 2, 568

\bibitem[{{Lou}(2001)}]{Lou2001}
{Lou}, Y.-Q. 2001, \apjl, 556, L121

\bibitem[{{Lynden-Bell} \& {Ostriker}(1967)}]{Lynden1967}
{Lynden-Bell}, D. \& {Ostriker}, J.~P. 1967, \mnras, 136, 293

\bibitem[{Paxton {et~al.}(2011)Paxton, Bildsten, Dotter, Herwig, Lesaffre, \&
  Timmes}]{Paxton2011}
Paxton, B., Bildsten, L., Dotter, A., {et~al.} 2011, The Astrophysical Journal
  Supplement Series, 192, 3

\bibitem[{{Perdang} \& {Blacher}(1982)}]{Perdang1982}
{Perdang}, J. \& {Blacher}, S. 1982, \aap, 112, 35

\bibitem[{{Provost} {et~al.}(2000){Provost}, {Berthomieu}, \&
  {Morel}}]{Provost2000}
{Provost}, J., {Berthomieu}, G., \& {Morel}, P. 2000, \aap, 353, 775

\bibitem[{{Ritzwoller} \& {Lavely}(1991)}]{Ritzwoller1991}
{Ritzwoller}, M.~H. \& {Lavely}, E.~M. 1991, \apj, 369, 557

\bibitem[{Schad(2013)}]{SchadPhD}
Schad, A. 2013, PhD thesis, Universit{\"a}t Freiburg im Breisgau

\bibitem[{Schad {et~al.}(2011)Schad, Timmer, \& Roth}]{Schad2011b}
Schad, A., Timmer, J., \& Roth, M. 2011, \apj, 734, 97

\bibitem[{{Schad} {et~al.}(2013){Schad}, {Timmer}, \& {Roth}}]{Schad2013}
{Schad}, A., {Timmer}, J., \& {Roth}, M. 2013, \apjl, 778, L38

\bibitem[{{Scherrer} \& {Gough}(2019)}]{ScherrerGough2019}
{Scherrer}, P.~H. \& {Gough}, D.~O. 2019, \apj, 877, 42

\bibitem[{{Schou} {et~al.}(1998){Schou}, {Antia}, {Basu}, {Bogart}, {Bush},
  {Chitre}, {Christensen-Dalsgaard}, {Di Mauro}, {Dziembowski}, {Eff-Darwich},
  {Gough}, {Haber}, {Hoeksema}, {Howe}, {Korzennik}, {Kosovichev}, {Larsen},
  {Pijpers}, {Scherrer}, {Sekii}, {Tarbell}, {Title}, {Thompson}, \&
  {Toomre}}]{Schou1998a}
{Schou}, J., {Antia}, H.~M., {Basu}, S., {et~al.} 1998, \apj, 505, 390

\bibitem[{{Schunker} {et~al.}(2018){Schunker}, {Schou}, {Gaulme}, \&
  {Gizon}}]{Schunker2018}
{Schunker}, H., {Schou}, J., {Gaulme}, P., \& {Gizon}, L. 2018, \solphys, 293,
  95

\bibitem[{{Thuras} {et~al.}(1935){Thuras}, {Jenkins}, \& {O'Neil}}]{Thuras1935}
{Thuras}, A.~L., {Jenkins}, R.~T., \& {O'Neil}, H.~T. 1935, The Journal of the
  Acoustical Society of America, 6, 173

\bibitem[{{Townsend} {et~al.}(2018){Townsend}, {Goldstein}, \&
  {Zweibel}}]{Townsend2018}
{Townsend}, R.~H.~D., {Goldstein}, J., \& {Zweibel}, E.~G. 2018, \mnras, 475,
  879

\bibitem[{{Townsend} \& {Teitler}(2013)}]{Townsend2013}
{Townsend}, R.~H.~D. \& {Teitler}, S.~A. 2013, \mnras, 435, 3406

\bibitem[{{Unno} {et~al.}(1989){Unno}, {Osaki}, {Ando}, {Saio}, \&
  {Shibahashi}}]{Unno1989}
{Unno}, W., {Osaki}, Y., {Ando}, H., {Saio}, H., \& {Shibahashi}, H. 1989,
  {Nonradial oscillations of stars}, 2nd edn. (Tokyo: University of Tokyo
  Press)

\bibitem[{{Weinberg} {et~al.}(2013){Weinberg}, {Arras}, \&
  {Burkart}}]{Weinberg2013}
{Weinberg}, N.~N., {Arras}, P., \& {Burkart}, J. 2013, \apj, 769, 121

\bibitem[{{Wentzel}(1987)}]{Wentzel1987}
{Wentzel}, D.~G. 1987, \apj, 319, 966

\bibitem[{{Westervelt}(1957)}]{Westervelt1957a}
{Westervelt}, P.~J. 1957, The Journal of the Acoustical Society of America, 29,
  199

\bibitem[{{Westervelt}(1963)}]{Westervelt1963}
{Westervelt}, P.~J. 1963, The Journal of the Acoustical Society of America, 35,
  535

\bibitem[{{Woodard}(1989)}]{Woodard1989}
{Woodard}, M.~F. 1989, \apj, 347, 1176

\bibitem[{{Woodard}(2007)}]{Woodard2007}
{Woodard}, M.~F. 2007, \apj, 668, 1189

\bibitem[{{Woodard}(2016)}]{Woodard2016}
{Woodard}, M.~F. 2016, \mnras, 460, 3292

\end{thebibliography}

%-------------------------------------------------------------------
% Appendix
%-------------------------------------------------------------------

\appendix

  \section{Wave operator}
\label{appWaveOp}

The Eulerian perturbations to the  solar structure variables are \citep[e.g.][]{JCDoscilAsBook,Basu2016}:
\begin{align}
p_1 &= -c_0^2 \rho_0 \nabla \cdot \bxi - \nabla p_0 \cdot \bxi, \label{eqp1}\\
\rho_1 &= - \nabla \cdot (\rho_0\bxi) \label{eqrho1} \\
\bg_1 &=  -\nabla \Phi_1 = -G\, \nabla_\br \int_V  \frac{\nabla_{\br'}\cdot(\rho_0(\br')\bxi(\br'))}{|\br-\br'|}\id^3\br'\label{eqg1}.
\end{align}
As a consequence, the wave operator $\cL$  in Eq.~\eqref{eqwaveOp} can be written as
\begin{align}
\cL[\bxi] &= \cL_1[\bxi] + \cL_2[\bxi] + \cL_3[\bxi] + \cL_4[\bxi] \label{eqcL}\\
\cL_1[\bxi] &=  -\frac{1}{\rho_0}  \nabla \Big( c_0^2 \rho_0 \nabla \cdot \bxi \Big) \\
&= - \nabla \Big( c_0^2 \nabla \cdot \bxi\Big)  -\frac{\nabla \rho_0}{\rho_0}  c_0^2  \nabla \cdot \bxi \\
\cL_2[\bxi] &= -\frac{1}{\rho_0}  \nabla \Big(\nabla p_0 \cdot \bxi  \Big) \\
\cL_3[\bxi] &=  \frac{1}{\rho_0} \nabla \cdot (\rho_0\bxi) \,\bg_0 = (\nabla \cdot \bxi) \,\bg_0  + \frac{1}{\rho_0} (\nabla\rho_0 \cdot\bxi) \,\bg_0 \\
\cL_4[\bxi] &=  G\, \nabla_\br \int_V \frac{\nabla_{\br'}\cdot(\rho_0(\br')\bxi(\br'))}{|\br-\br'|}\id^3\br' \label{eqcL4}.
\end{align}

\section{Checking the eigenfunction computation}
\label{appEF}

To check the correctness of the eigenfunction computations, we have performed a number of tests. 

\begin{figure}
	\centering

	\includegraphics[width=1\linewidth]{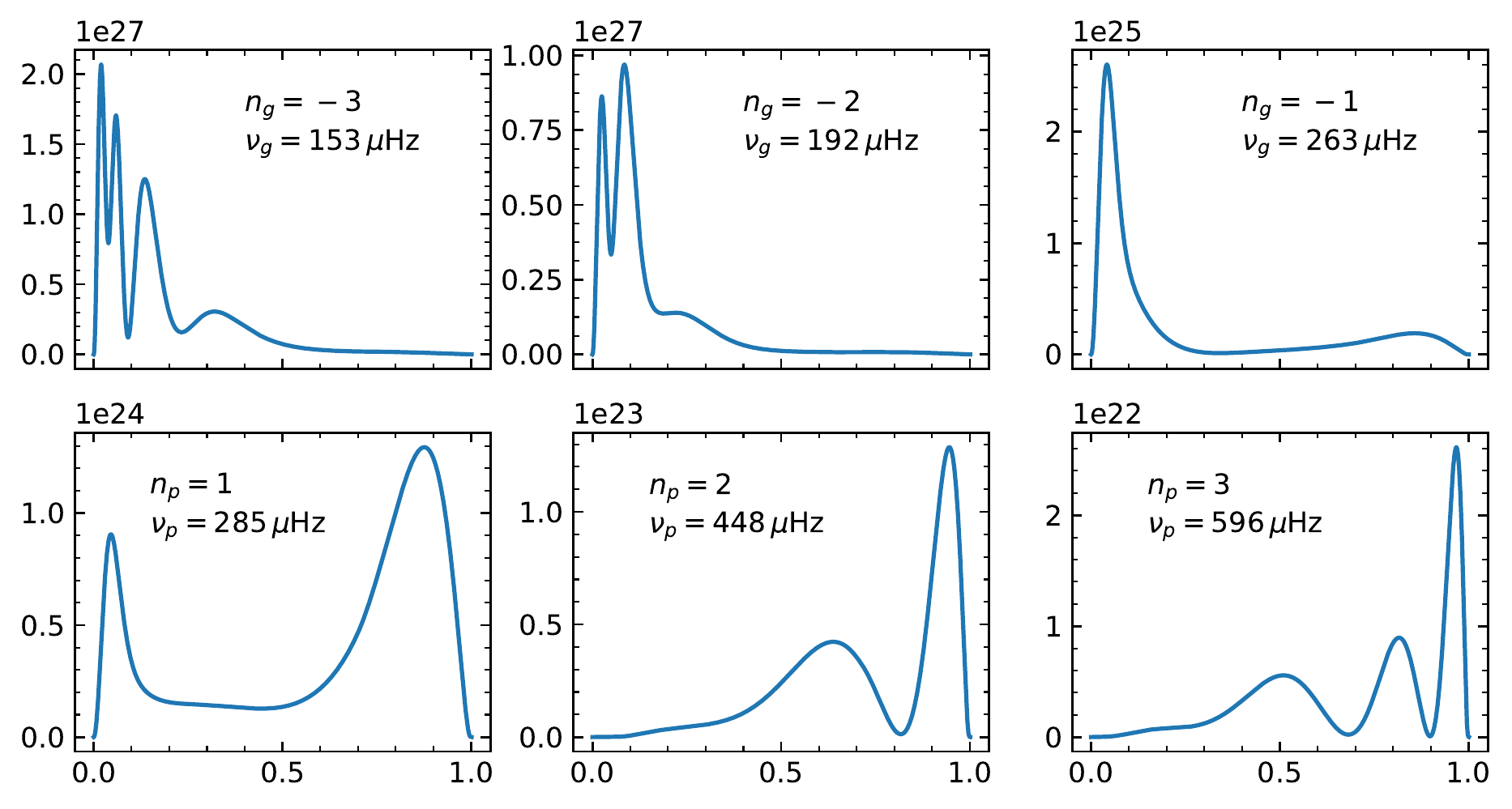}
	\caption{Kinetic energy density ($\rho r^2 (\xi_{r}(r)^2 + l(l+1)\xi_{h}(r)^2)$ $[\rm{g}\,\rm{cm}]$) as a function of fractional solar radius $r/R_\sun$ for modes with degree $l_p=l_g=1$. The radial orders and frequencies of the modes are given in each panel. Compare to Figure~1 from \cite{Provost2000}. We note that the mode frequencies are slightly different compared to \cite{Provost2000} due to slight differences in the solar model.}
	\label{figprovost2000fig1}
\end{figure}

\begin{figure}
	\centering

	\includegraphics[width=1\linewidth]{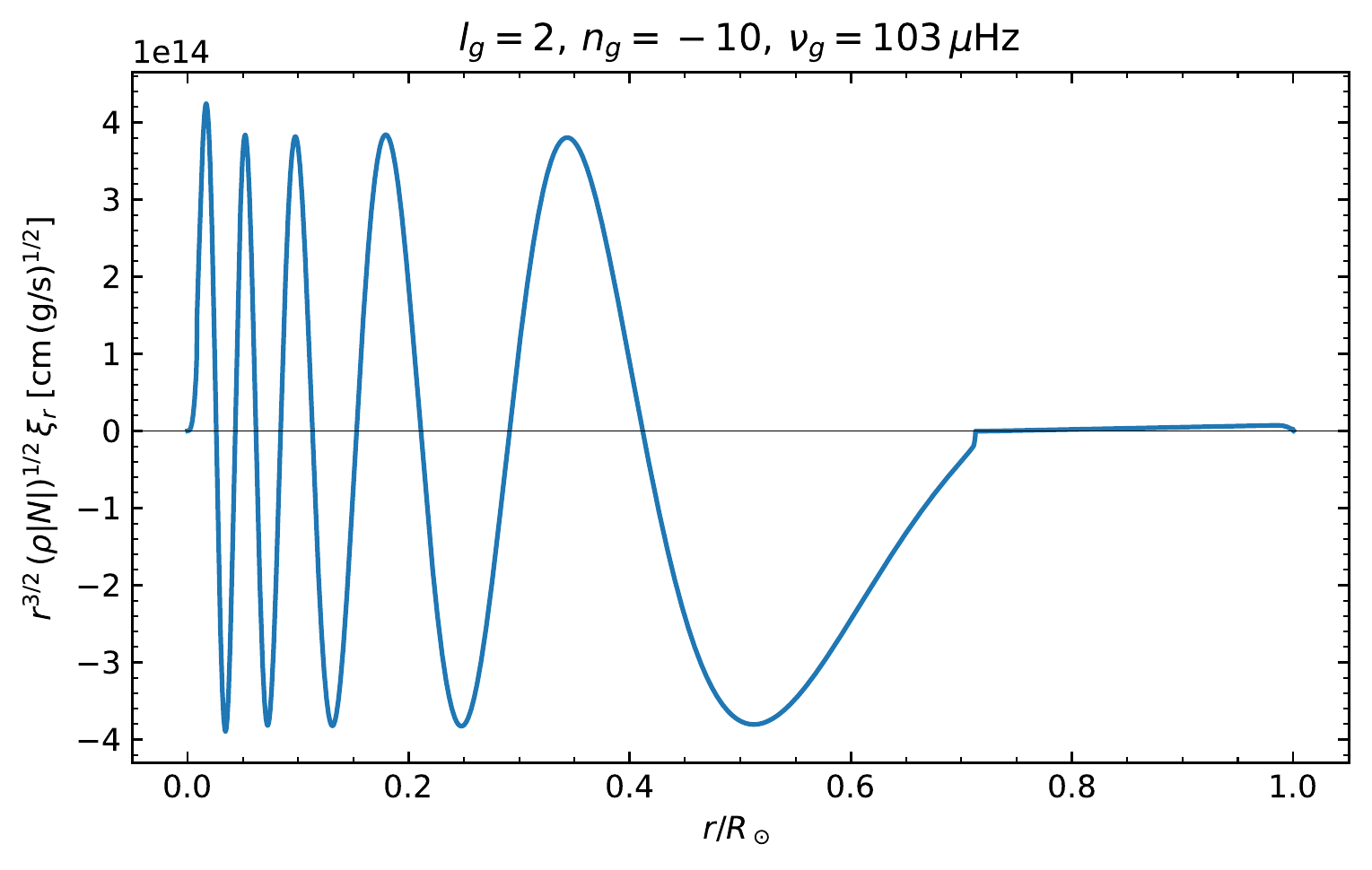}

	\caption{Scaled radial g-mode eigenfunction. Compare to Figure~7.4 in \cite{JCDoscilAsBook}.}
	\label{figJCDfig7_4}
\end{figure}

\begin{figure}
	\centering

	\includegraphics[width=1\linewidth]{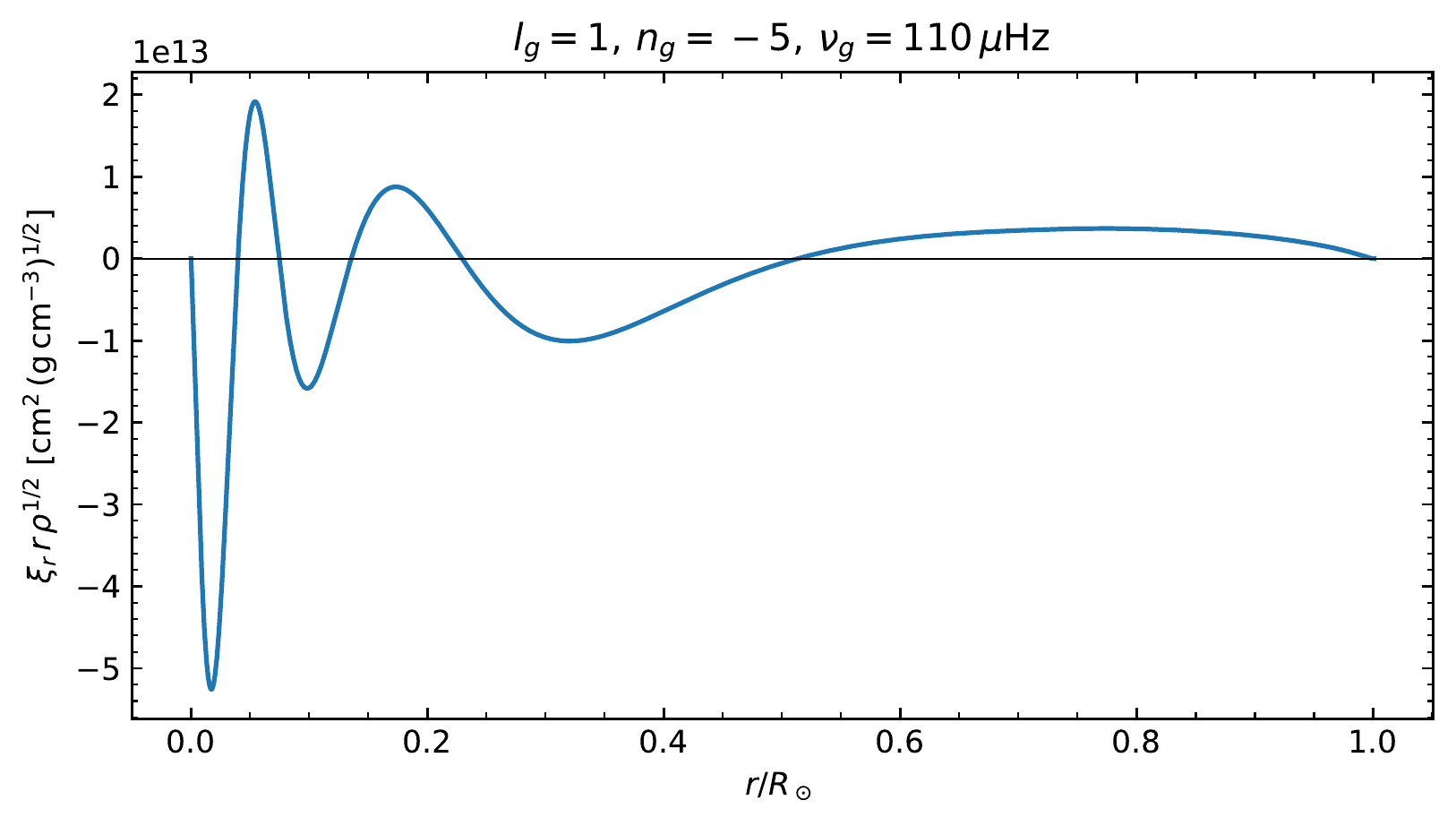}

	\caption{Scaled radial g-mode eigenfunction. Compare to panel a) in Figure~5.10 from \cite{JCDoscilAsBook}.}
	\label{figJCDfig5_10}
\end{figure}

Firstly, we reproduced plots of eigenfunctions in the literature, namely \citet[Figs.~6, 7a, 7b]{Berthomieu1990}, \citet[Fig.~1]{Provost2000}, and \citet[Figs.~5.10 and~7.4]{JCDoscilAsBook} with the results from our eigenfunction computation, see Figures~\ref{figprovost2000fig1} -~\ref{figJCDfig5_10}. The results look the same as in the literature.

\begin{figure}
	\centering

	\includegraphics[width=1\linewidth]{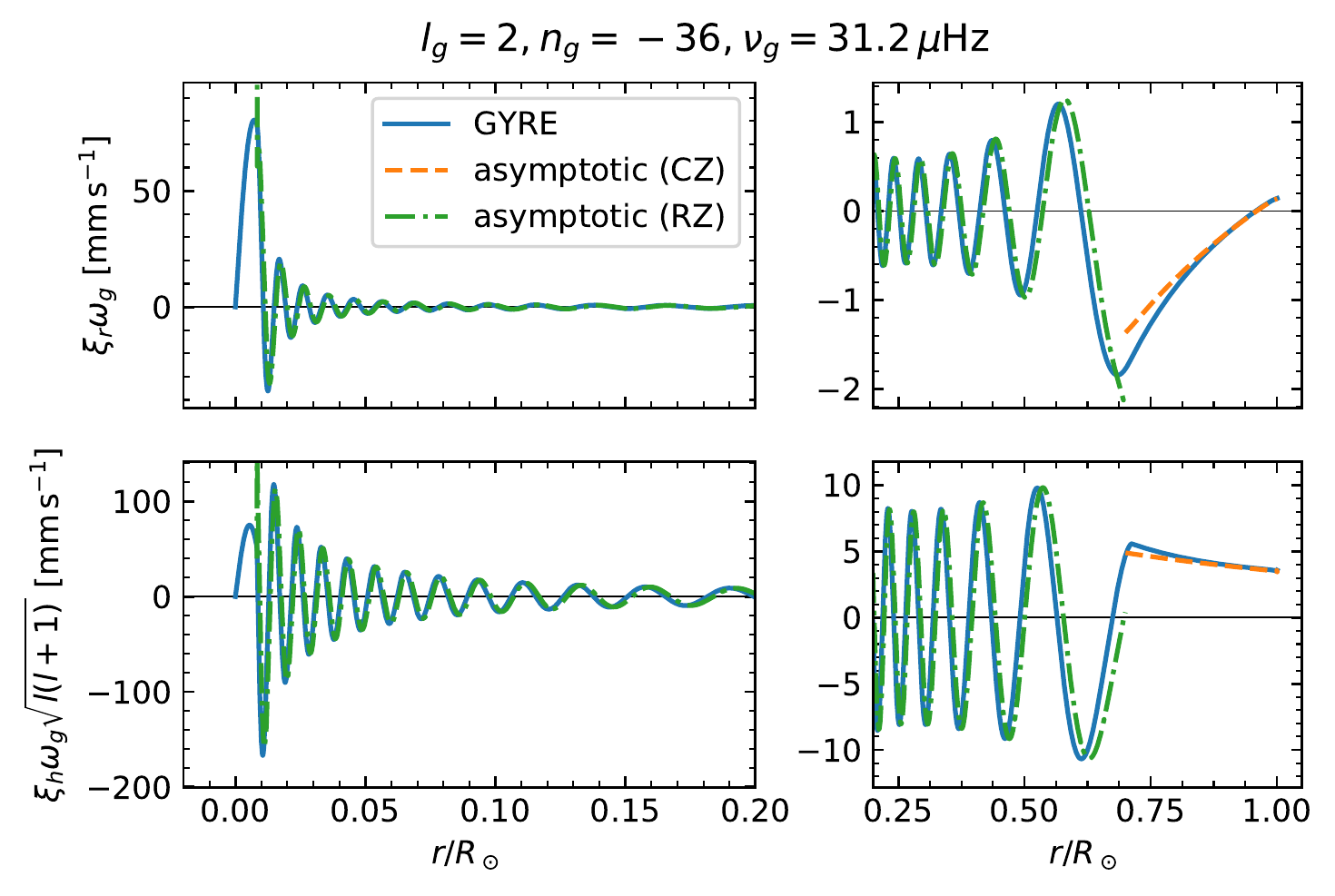}

	\caption{Comparison of radial (top) and horizontal (bottom) g-mode eigenfunctions (computed with \GYRE) with asymptotic formulae. For the convection zone (CZ) and radiative zone (RZ), different approximations were used, see Appendix~\ref{appEF}. The same mode as in Figure~\ref{figExampleEulg36} is shown.}
	\label{figAsymptEf}
\end{figure}

\begin{figure}
	\centering

	\includegraphics[width=1\linewidth]{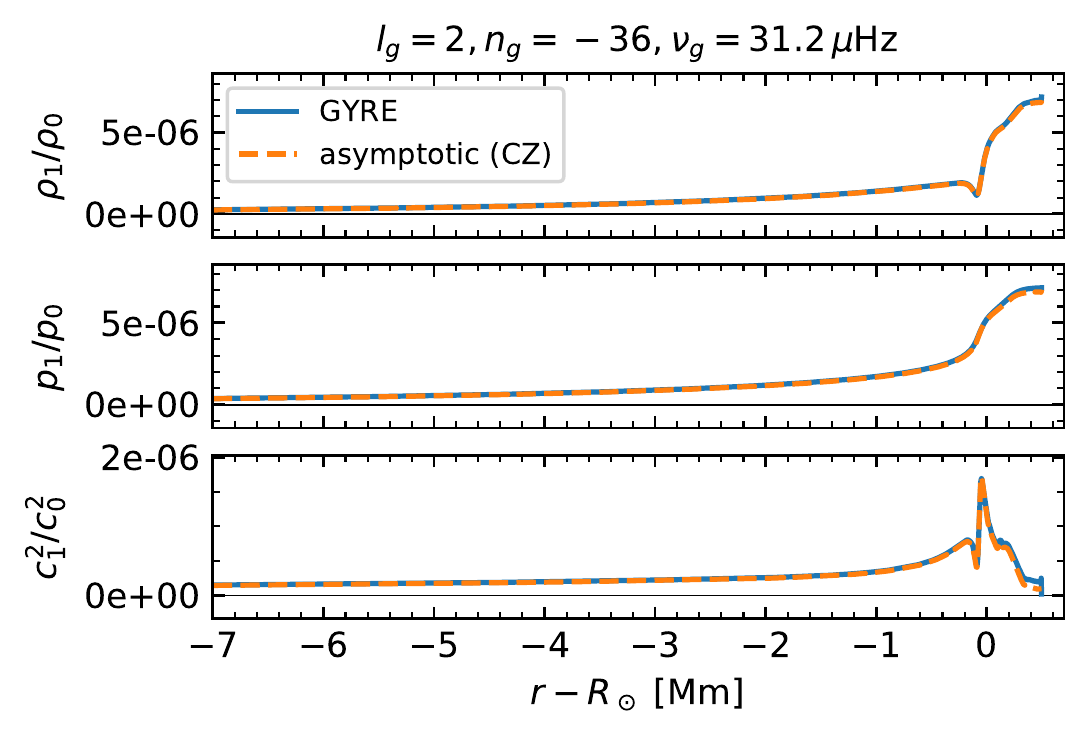}

	\caption{Comparison of g-mode solar structure eigenfunctions (computed with \GYRE) with asymptotic formulae for the convection zone (CZ). The same mode as in Figure~\ref{figExampleEulg36} is shown. See Appendix~\ref{appEF} for details.}
	\label{figAsymptEfStruct}
\end{figure}

Secondly, we compared our results for g-mode eigenfunctions to asymptotic relations, both for the trapping region \citep[Eqs.~7.129 and~7.131 in][]{JCDoscilAsBook} and the evanescent region, which mostly coincides with the convection zone \citep[Eqs.~7.137 and~7.142 in][]{JCDoscilAsBook}. We find that the asymptotic relations qualitatively well reproduce the {displacement} eigenfunctions, and in most cases, a reasonable approximation to the {displacement} eigenfunction can be obtained by glueing together eigenfunctions obtained from the asymptotic relations for the trapping and evanescent regions near the bottom of the convection zone, see Figure~\ref{figAsymptEf}. From the asymptotic displacement eigenfunctions, asymptotic expressions for the structure eigenfunctions can be obtained using Eqs.~\eqref{eqp1}, \eqref{eqrho1}, and \eqref{eqdc1tilde}, which agree well with the full structure eigenfunctions {near the surface}, see Figure~\ref{figAsymptEfStruct}.

Thirdly, we checked for modes of low order that eigenfunctions obtained using \GYRE~and \ADIPLS~agree closely provided the same solar model and the same boundary conditions are used.

\section{The perturbed wave operator without hydrostatic equilibrium}
\label{appPerturbedWaveOp}

In the following, we compute the perturbation to the wave operator $\delta \cL$ in the presence of perturbations to solar structure, $\delta\rho_0,\delta p_0,\delta(c_0^2),\delta \bg_0$, which potentially do not fulfil the hydrostatic equilibrium condition $\delta \nabla p_0 = \delta(\rho_0 \bg_0)$. Omitting the zero subscript in perturbed quantities, for example $\delta\rho=\delta\rho_0$, and
perturbing Equations~\eqref{eqcL}-\eqref{eqcL4}, we obtain
\begin{align}
\delta \cL_1[\bxi] &=  - \nabla \Big( \delta(c^2) \nabla \cdot \bxi\Big)  - \frac{\nabla \delta\rho}{\rho_0}  c_0^2  \nabla \cdot \bxi  \nonumber\\
&\quad +   \frac{\nabla \rho_0}{\rho_0^2} \delta \rho \,\, c_0^2  \nabla \cdot \bxi -  \frac{\nabla \rho_0}{\rho_0} \delta(c^2)   \nabla \cdot \bxi \label{eqdL1}
\end{align}
\begin{align}
\delta \cL_2[\bxi] &=  \frac{\delta \rho}{\rho_0^2}\nabla \Big( \nabla p_0 \cdot \bxi \Big) - \frac{1}{\rho_0}\nabla \Big( \nabla \delta p \cdot \bxi \Big) 
\end{align}
\begin{align}
\delta \cL_3[\bxi] &= (\nabla \cdot \bxi) \,\delta\bg  - \frac{\delta\rho}{\rho_0^2} (\nabla\rho_0 \cdot\bxi) \,\bg_0 \nonumber\\
&\quad+  \frac{1}{\rho_0} (\nabla \delta\rho \cdot\bxi) \,\bg_0  +  \frac{1}{\rho_0} (\nabla\rho_0 \cdot\bxi) \,\delta \bg \label{eqdL3}
\end{align}
\begin{align}
\delta\cL_4[\bxi] &=  G \, \nabla_\br \int_V \frac{\nabla_{\br'}\cdot(\delta\rho(\br')\bxi(\br'))}{|\br-\br'|}\id^3\br' \label{eqdL4} 
\end{align}

In addition, in the presence of a {perturbation to the} flow field $\delta\bu(\br)$, this flow gives rise to a fifth perturbation term \citep[e.g.][]{JCDoscilAsBook}
\begin{align}
\delta \cL_5 [\bm{\xi}] &= -2i\omega({\delta}\bu \cdot \nabla)\bm{\xi} \label{eqdL5},
\end{align}
such that $\delta \cL= \sum_{i=1}^5 \delta \cL_i$.

\section{Perturbation to p-mode frequencies}
\label{appPerturb}

We will now assume that the perturbation to solar structure is given by a g mode as outlined in Eqs.~\eqref{eqdp1}-\eqref{equ}. Given these perturbations, we will determine the perturbation to p-mode frequencies using Eq.~\eqref{eqFreqshiftRotBackground3} and Eqs.~ \eqref{eqdL1}-\eqref{eqdL5}.

{For brevity in notation, we use
\begin{align}
(l,m,n)&=(l_p,m_p,n_p), \\
(l',m',n')&= (l_g,m_g,n_g)
\end{align}
in Appendices~\ref{appPerturb} and ~\ref{appHorizInt}.}

\subsection{Some useful relations}
\label{appuseful}

To ease the necessary derivations, we assume that in addition to the eigenfunctions $\xi_{r,g},\xi_{h,g}$, the Eulerian perturbations $\pOneg,\rhoOneg,\PhiOneg$ are given and we separate their radial and horizontal dependence \citep[e.g.][]{Aerts2010},
\begin{align}
\pOneg &= \pOnegt(r) Y_{l'm'}, \label{eqp1gtilde}\\
\rhoOneg &= \rhoOnegt(r) Y_{l'm'}, \label{eqrho1gtilde}\\
\PhiOneg &= \PhiOnegt(r) Y_{l'm'}. \label{eqPhi1gtilde}
\end{align}
Similarly, we obtain using Eqs.~(\ref{eqdp1}~-~\ref{eqdGamma}) and Eqs.~(\ref{eqp1gtilde}~-~\ref{eqrho1gtilde}),
\begin{align}
\delta(c^2) &= \cSqOnegt(r) Y_{l'm'},\\
\cSqOnegt &= -\frac{\Gamma_{1,0} \,p_0}{\rho_0^2} \rhoOnegt  + \frac{\Gamma_{1,0}}{\rho_0} \pOnegt - \frac{p_0}{\rho_0} \dv{\Gamma_{1,0}}{r} \xi_{r,g}. \label{eqdc1tilde}
\end{align}
In addition, we need
\begin{align}
\bgOneg &= -\nabla \PhiOneg = -\nabla (\PhiOnegt Y_{l'm'}) \\
&= -\dv{\PhiOnegt}{r} Y_{l'm'} \unitr - \frac{\PhiOnegt}{r} \bPsi_{l'm'}
\end{align}
and an expression for the divergence of the displacement eigenfunction,
\begin{align}
\nabla\cdot\bxi_p &= \nabla\cdot\left(\xi_{r,p} Y_{lm} \unitr \right) +\nabla\cdot\left(\xi_{h,p} \bPsi_{lm} \right) \\
&= \left(\dv{\xi_{r,p}}{r} + \frac{2}{r} \xi_{r,p} - \frac{l(l+1)}{r} \xi_{h,p} \right) Y_{lm} \label{eqDivXiDef} \\
&= \divxip(r) Y_{lm} \label{eqdivxi},
\end{align}
where the function $\divxip(r)$ is defined as the term in the large brackets in Eq.~\eqref{eqDivXiDef}.

\subsection{Solar structure terms}

We first separately look at the perturbations to solar structure, which means contributions to $\delta(\omega_p^2)$ arising from $\delta\cL_1$ till $\delta\cL_{3}$. We omit one term ($\delta\cL_4$) for simplicity. This term is expected to have a small contribution to the wave equation. It models the change in Eulerian perturbation to the gravitational potential $\Phi_{1,p}$ due to the presence of the g mode. To simplify our derivations, {we write $\delta\cL_{i-j}$ for the $j$th term in $\delta\cL_i$ in Eqs.~\eqref{eqdL1}-\eqref{eqdL3} and we use the relations given in Appendix~\ref{appuseful}.}

We start by the first term in $\delta\cL_1$ (Eq.~[\ref{eqdL1}]),
\begin{align} 
% L 1-1
\delta \mathcal{L}_{1-1} [\bxi_p] &= -\nabla \left(\delta(c^2) \nabla \cdot\bxi_p \right) = - \nabla \left( \cSqOnegt Y_{l'm'} \,\divxip Y_{lm} \right) \label{eqL11eval}\\
&= - \left( \dv{\cSqOnegt}{r} \,  Y_{l'm'} \bm{\hat r}  + \frac{\cSqOnegt}{r}\bm\Psi_{l'm'} \right) \,\divxip Y_{lm} \nonumber \\
&\quad - \cSqOnegt Y_{l'm'} \left( \dv{\divxip}{r} Y_{lm} \bm{\hat r} + \frac{\divxip}{r}\bm\Psi_{lm} \right).
\end{align}
The frequency shift arising from this term can then be evaluated using Eq.~\eqref{eqFreqshiftRotBackground3} and Eq.~\eqref{eqxiYlm} applied to the p-mode eigenfunction,
\begin{align} 
% Integral of L 1-1 
\langle \bxi_p, \delta \mathcal{L}_{1-1} [\bxi_p] \rangle 
&= -\int \xi_{r,p} \left(\dv{\cSqOnegt}{r} \,\divxip + \cSqOnegt \dv{\divxip}{r} \right) \rho_0 r^2 \,\id{r} \nonumber \\
&\quad\quad\times \int Y_{lm}^* Y_{l'm'} Y_{lm} \,\id{\Omega} \nonumber\\
&\quad- \int \xi_{h,p}\, \cSqOnegt \divxip \, \rho_0 r \,\id{r} \,\,\Bigg(\int \bm\Psi_{lm}^* \cdot \bm\Psi_{l'm'} Y_{lm} \,\id{\Omega} \nonumber \\
&\quad\quad + \int  \bm\Psi_{lm}^* Y_{l'm'} \cdot \bm\Psi_{lm}  \,\id{\Omega} \Bigg).
\label{eqint1-1}
\end{align}
{Here, $\id\Omega=\sin\theta\,\id\theta\,\id\phi$ is the surface element on the unit sphere.}

For the following terms, we proceed in a similar way and obtain
\begin{align}
% L 1-2 
\delta \mathcal{L}_{1-2} [\bxi_p] &= - \frac{\nabla \delta \rho}{\rho_0} c_0^2\nabla \cdot \bxi_p = -\frac{c_0^2}{\rho_0} \nabla (\rhoOnegt Y_{l'm'}) \, \nabla \cdot \bxi_p \\
&= -\frac{c_0^2}{\rho_0} \left(\dv{\rhoOnegt}{r} \, Y_{l'm'} \bm{\hat r} + \rhoOnegt \frac{1}{r}\bm{\Psi}_{l'm'} \right) \,\divxip \,Y_{lm}, \\
%% Integral of L 1-2
\langle \bxi_p, \delta \mathcal{L}_{1-2} [\bxi_p] \rangle
&= -\int \xi_{r,p} c_0^2 \dv{\rhoOnegt}{r}\,\divxip r^2 \,\id{r} \int Y_{lm}^* Y_{l'm'} Y_{lm} \,\id{\Omega} \nonumber \\
&\quad -\int \xi_{h,p} c_0^2 \, \rhoOnegt \,\divxip r \,\id{r} \int \bm\Psi_{lm}^* \cdot \bm\Psi_{l'm'} Y_{lm} \,\id{\Omega} \,.
\label{eqint1-2}
\end{align}
Using similar intermediate steps, we obtain the results for the following terms,
\begin{align}
% L 1-3
%% Integral of L 1-3
\langle \bxi_p, \delta \mathcal{L}_{1-3} [\bxi_p] \rangle
&= \int \xi_{r,p} \frac{\rhoOnegt}{\rho_0} \dv{\rho_0}{r}\,  c_0^2 \,\divxip r^2 \,\id{r} \int Y_{lm}^* Y_{l'm'} Y_{lm} \,\id{\Omega}, \label{eqint1-3} \\
% L 1-4
%% Integral of L 1-4
\langle \bxi_p, \delta \mathcal{L}_{1-4} [\bxi_p] \rangle 
&= - \int \xi_{r,p} \dv{\rho_0}{r}\, \cSqOnegt \,\divxip r^2\,\id{r} \int Y_{lm}^* Y_{l'm'} Y_{lm} \,\id{\Omega}, \label{eq:int1-4}
\\
% L 2-1
% Integral of L 2-1
\langle \bxi_p, \delta \mathcal{L}_{2-1} [\bxi_p] \rangle 
&= \int \xi_{r,p} \frac{1}{\rho_0} \rhoOnegt \left(\dv[2]{p_0}{r} \xi_{r,p} + \dv{p_0}{r} \dv{\xi_{r,p}}{r} \right) r^2\,\id{r} \nonumber \\
&\quad\quad\times\int Y_{lm}^* Y_{l'm'} Y_{lm} \,\id{\Omega} \label{eq:int2-11} \\
&\quad + \int \xi_{h,p} \frac{1}{\rho_0} \rhoOnegt \dv{p_0}{r} \xi_{r,p} r \,\id{r} \nonumber \\
&\quad\quad\times
\int \bm{\Psi}_{lm}^*  Y_{l'm'} \cdot \bm{\Psi}_{lm} \,\id{\Omega} \,.
\label{eq:int2-12}
\end{align}

The next term requires a bit more work. We start with
\begin{align}
\delta \mathcal{L}_{2-2} [\bxi_p] &= -\frac{1}{\rho_0} \nabla \left(\dv{\pOnegt}{r} Y_{l'm'}\xi_{r,p} Y_{lm} + \pOnegt\frac{1}{r} \xi_{h,p} \bm{\Psi}_{l'm'} \cdot \bm{\Psi}_{lm} \right),\label{eq:dL2-2} \\
&= -\frac{1}{\rho_0} \Bigg( \nabla\bigg(\dv{\pOnegt}{r} \xi_{r,p}\bigg) Y_{l'm'} Y_{lm} + \dv{\pOnegt}{r} \xi_{r,p} \nabla\bigg(Y_{l'm'}Y_{lm}\bigg) \\
&\quad + \nabla\bigg(\pOnegt\frac{1}{r} \xi_{h,p} \bigg)\bm{\Psi}_{l'm'} \cdot \bm{\Psi}_{lm} \nonumber \\
&\quad + \pOnegt\frac{1}{r} \xi_{h,p} \nabla\bigg(\bm{\Psi}_{l'm'} \cdot \bm{\Psi}_{lm}\bigg)\Bigg),\\
&= -\frac{1}{\rho_0} \Bigg( \bigg(\dv{^2\pOnegt}{r^2} \xi_{r,p} + \dv{\pOnegt}{r} \dv{\xi_{r,p}}{r}\bigg) Y_{l'm'} Y_{lm} \, \unitr \\
&\quad \quad + \dv{\pOnegt}{r} \xi_{r,p} \,\frac{1}{r}\bigg(\bPsi_{l'm'}Y_{lm}+Y_{l'm'}\bPsi_{lm}\bigg) \\
&\quad \quad+ \bigg(\dv{\pOnegt}{r}\frac{1}{r} \xi_{h,p} -\pOnegt\frac{1}{r^2} \xi_{h,p} + \pOnegt\frac{1}{r} \dv{\xi_{h,p}}{r}\bigg) \nonumber \\
&\quad\quad\quad\times \bigg(\bm{\Psi}_{l'm'} \cdot \bm{\Psi}_{lm}\bigg) \, \unitr \\
& \quad \quad + \pOnegt\frac{1}{r} \xi_{h,p} \, \nabla\bigg(\bm{\Psi}_{l'm'} \cdot \bm{\Psi}_{lm}\bigg)\Bigg), \label{eqdL22last}
\end{align}
To evaluate the last term in $\langle \bxi_p, \delta \mathcal{L}_{2-2} [\bxi_p] \rangle$ arising from Eq.~\eqref{eqdL22last}, we first note that $\nabla(\bm{\Psi}_{l'm'} \cdot \bm{\Psi}_{lm})$ is purely horizontal. The horizontal gradient $\nabla_h= \unitth \frac{\partial}{\partial \theta} + \unitph \frac{1}{\sin\theta} \frac{\partial}{\partial \phi}$ applied to a function $f(\theta,\phi)$ that is independent of radius, satisfies $r\nabla f  = \nabla_h f$. It can then be shown using Gauss's divergence theorem and the property $\nabla \cdot (g(r) \bPsi_{lm})=-\frac{l(l+1)}{r}g(r)Y_{lm}$ that
\begin{align}
\int &\bPsi_{lm}^* \cdot \nabla_h (\bPsi_{l'm'} \cdot \bPsi_{lm}) \, \id \Omega = l(l+1) \int Y_{lm}^* (\bPsi_{l'm'} \cdot \bPsi_{lm})   \,\id \Omega.
\end{align}
Using this, we can now evaluate
\begin{align}
\langle \bxi_p, \delta \mathcal{L}_{2-2} [\bxi_p] \rangle &= 
 -\int r^2 \xi_{r,p} \left(\dv{^2\pOnegt}{r^2} \xi_{r,p} + \dv{\pOnegt}{r} \dv{\xi_{r,p}}{r}\right) \id r \nonumber\\
 & \quad\quad \times\int Y_{lm}^* Y_{l'm'} Y_{lm} \, \id\Omega \nonumber\\
&\quad   -\int r^2 \xi_{h,p} \dv{\pOnegt}{r} \xi_{r,p} \,\frac{1}{r} \id r \nonumber\\
& \quad \quad \times \int \bPsi_{lm}^* \cdot \left(\bPsi_{l'm'}Y_{lm}+Y_{l'm'}\bPsi_{lm}\right) \,\id \Omega\nonumber\\
&\quad -\int \left(\dv{\pOnegt}{r}\frac{1}{r} \xi_{h,p} -\pOnegt\frac{1}{r^2} \xi_{h,p} + \pOnegt\frac{1}{r} \dv{\xi_{h,p}}{r}\right) \nonumber \\
&\quad\quad\quad \times r^2 \xi_{r,p} \,\id r \, 
\int Y_{lm}^* \,(\bm{\Psi}_{l'm'} \cdot \bm{\Psi}_{lm}) \, \id\Omega \nonumber\\
&  \quad -\int r^2 \xi_{h,p} \pOnegt\frac{1}{r^2} \xi_{h,p} \,\id r \nonumber \\
& \quad \quad \times l(l+1)\int Y_{lm}^* \,(\bm{\Psi}_{l'm'} \cdot \bm{\Psi}_{lm})\,\id\Omega.
\label{eqint2-2b}
\end{align}

The remaining terms can again be evaluated in a similar way as in Eqs.~(\ref{eqL11eval}~-~\ref{eqint1-1}),
%%%  L_3
\begin{align}
% L3-1
% Integral of L3-1
\langle \bxi_p, \delta \mathcal{L}_{3-1} [\bxi_p]\rangle &= -\int \xi_{r,p} \dv{\PhiOnegt}{r} \divxip \, \rho_0 r^2 \, \id{r} \int Y_{lm}^* Y_{l'm'} Y_{lm} \, \id{\Omega} \notag\\
&\quad - \int \xi_{h,p} \PhiOnegt \,\divxip \, \rho_0 r \,\id{r} \nonumber \\
&\quad\quad\times\int \bm{\Psi}_{lm}^* \cdot \bm{\Psi}_{l'm'} Y_{lm} \,\id{\Omega},
\label{eq:int3-1}
\end{align}
\begin{align}
% L3-2
% Integral of L3-2
\langle \bxi_p, \delta \mathcal{L}_{3-2} [\bxi_p]\rangle &= -\int (\xi_{r,p})^2 \frac{1}{\rho_0^2} \rhoOnegt \dv{\rho_0}{r} \dv{p_0}{r} r^2 \,\id{r} \nonumber \\
&\quad\quad\times\int Y_{lm}^* Y_{l'm'} Y_{lm} \,\id{\Omega},
\label{eq:int3-2}
\end{align}
\begin{align}
% L3-3
% Integral of L3-3
\langle \bxi_p, \delta \mathcal{L}_{3-3} [\bxi_p]\rangle &= \int (\xi_{r,p})^2 \dv{\rhoOnegt}{r} \frac{1}{\rho_0} \dv{p_0}{r} r^2 \,\id{r} \int Y_{lm}^* Y_{l'm'} Y_{lm} \,\id{\Omega} \notag\\
&\quad + \int \xi_{r,p} \, \rhoOnegt \, \xi_{h,p} \frac{1}{\rho_0} \dv{p_0}{r} r \,\id{r} \nonumber \\
&\quad\quad\times\int Y_{lm}^* \bm{\Psi}_{l'm'} \cdot \bm{\Psi}_{lm} \,\id{\Omega},
\label{eq:int3-3}
\end{align}
\begin{align}
% L3-4
% Integral of L3-4
\langle \bxi_p, \delta \mathcal{L}_{3-4} [\bxi_p]\rangle &= -\int (\xi_{r,p})^2 \dv{\rho_0}{r} \dv{\PhiOnegt}{r}  r^2 \,\id{r} \int Y_{lm}^* Y_{l'm'} Y_{lm} \,\id{\Omega} \notag\\
&\quad - \int \xi_{h,p} \dv{\rho_0}{r} \xi_{r,p} \PhiOnegt r \,\id{r} \int \bm{\Psi}_{lm}^* \cdot \bm{\Psi}_{l'm'} Y_{lm} \,\id{\Omega} . \label{eq:int3-4}
\end{align}

\subsection{Flow terms}
\label{appFlow}

We now compute the terms due to the flow field caused by the g-mode that are {involved in} $\delta\cL_5$. The perturbation to the solar model due to flows is given by Eq.~\eqref{equ}, so that, from Eq.~\eqref{eqdL5},
\begin{align}
\delta \cL_5 [\bxi_p] &= - 2\omega_p \omega_g (\bxi_g \cdot \nabla)\bxi_p.\label{eqdcL5mat}
\end{align}
The computation of $\langle \bxi_p, \delta \cL_5 [\bxi_p]\rangle$ has essentially been done by \citet{Lavely1992}. This is because a g mode produces a poloidal flow field which can be expanded in vector spherical harmonics as \citep[][ Eqs.~1-3, C14]{Lavely1992}
\begin{align}
    \bu(\br) &= \sum_{s,t} u_{st}(r) Y_{st} \unitr + v_{st}(r) \bPsi_{st}, \\
    u_{st}(r) &= - \delta_{s,l'} \delta_{t,m'} \, \ii \omega_g \, \xi_{r,g}(r), \\
    v_{st}(r) &= - \delta_{s,l'} \delta_{t,m'}\,\ii \omega_g \, \xi_{h,g}(r). \\
\end{align}
From Eqs.~C31 till~C33 in \citet{Lavely1992}, we then obtain,
\begin{align}
\langle \bxi_p, \delta\cL_5 [\bxi_p] \rangle  &= -2\ii\omega_p  \int \bxi_{p}^* \cdot (\bu \cdot \nabla) \bxi_p \, \rho_0 \id^3{\br} \\
&= - 2\ii\omega_p \, 4\pi\,(-1)^m  \, \gamma_{l}\gamma_{l} \gamma_{l^{'}} \begin{pmatrix} l & l' & l \\ -m & m' & m \end{pmatrix} \nonumber\\
& \quad\quad \times\int \rho_0 [u_{l'm'} \tilde R_{l'} + v_{l'm'} \tilde H_{l'} ] r^2\,\id r \\
&= - 2\omega_p \omega_g \, 4\pi\,(-1)^m \, \gamma_{l}\gamma_{l} \gamma_{l^{'}} \begin{pmatrix} l & l' & l \\ -m & m' & m \end{pmatrix} \label{eqFlowTermGeneral}\\
& \quad\quad \times\int \rho_0 [\xi_{r,g} \tilde R_{l'} + \xi_{h,g} \tilde H_{l'} ] r^2\,\id r, \nonumber
\end{align}
where the Wigner 3j symbols are given for instance in \citet{Dahlen1998GlobalSeismology}, and where
\begin{align}
\gamma_l &= \sqrt{\frac{2l+1}{4\pi}} \\
\tilde R_{l'} &= \xi_{r,p} \dv{\xi_{r,p}}{r} B_{ll'l}^{(0)+} + \xi_{h,p} \dv{\xi_{h,p}}{r} B_{ll'l}^{(1)+} \\
r \tilde H_{l'} &= [\xi_{r,p}^2 - \xi_{r,p} \xi_{h,p}] B_{l'll}^{(1)+} \nonumber  \\
& \quad + \xi_{h,p}^2 \, (1 + (-1)^{2l+l'}) \Omega_0^l \Omega_0^l \Omega_0^{l'} \Omega_2^l  \begin{pmatrix} l & l' & l \\ 1 & 1 & -2 \end{pmatrix} \nonumber\\
& \quad + [\xi_{h,p} \xi_{r,p} - \Omega_0^l \Omega_0^l \xi_{h,p}^2] B_{l'll}^{(1)+}\\
\Omega_N^l &= \sqrt{\frac{(l+N)(l-N+1)}{2}} \\
B_{l'l''l}^{(N)\pm} &= \frac{1}{2} (1\pm(-1)^{l'+l''+l})\left[ \frac{(l'+N)!(l+N)!}{(l'-N)!(l-N)!} \right]^{\frac{1}{2}} \nonumber \\
&\quad \times(-1)^N \begin{pmatrix} l' & l'' & l \\ -N & 0 & N \end{pmatrix}. \label{eqBlllDef}
\end{align}

\section{Horizontal integrals and selection rules}
\label{appHorizInt}

We will now evaluate the horizontal integrals appearing in Appendix~\ref{appPerturb} and identify selection rules that must be satisfied for a first-order frequency shift to be present.

We first start by observing that the integration over $\phi$, which is part of all horizontal integrals, can always be written as
\begin{align}
\int_0^{2\pi} e^{-im\phi} e^{im'\phi} e^{im\phi} \,\id{\phi} 
&= 2\pi \delta_{m', 0} \,.
\label{eq:phi}
\end{align}
This integral vanishes when $m'=m_g \neq 0$. We therefore conclude that a first-order frequency shift is only present if the g mode has azimuthal degree $m_g=0$.

We now start with the simplest among the horizontal integrals, which can be evaluated using Gaunt's formula, see Equation \eqref{eq:3sph}, 
\begin{align}
\int Y_{lm}^* Y_{l'm'} Y_{lm} \,\id{\Omega} &= {4\pi}\,(-1)^m {\gamma_l \gamma_l \gamma_{l'}}
\begin{pmatrix} l & l' & l \\ 0 & 0 & 0 \end{pmatrix}
\begin{pmatrix} l & l' & l \\ -m & m' & m \end{pmatrix}.\label{eq:angular1}
\end{align}
From the selection rules of the Wigner 3-j symbols, we can deduct selection rules for this angular integral, which are summarised in Section~\ref{secSelectionRules}. As we will see, the remaining angular integrals either satisfy the same selection rules or evaluate to zero.

Next we consider the angular integral $\int Y_{lm}^* \bPsi_{l'm'} \cdot \bPsi_{lm} \,\id{\Omega}$. Using generalised spherical harmonics $Y_{lm}^\pm = Y_{lm}^{\pm 1}$, for which a definition is given in Appendix~\ref{appGaunt}, the vector spherical harmonic $\bPsi_{lm}$ can be written as, see Eq.~(C.139) in \citet{Dahlen1998GlobalSeismology},
\begin{equation}
\bPsi_{lm} = r {\nabla}_h Y_{lm} = {\Omega_0^l} \, (Y_{lm}^- \bm{\hat e_-} + Y_{lm}^+ \bm{\hat e_+}),
\end{equation}
where {the vectors $\bm{\hat e_\pm}$} are defined as, see Eq. (C.49) in \cite{Dahlen1998GlobalSeismology},
\begin{equation}
\bm{\hat e_-} = \frac{1}{\sqrt{2}}(\bm{\hat \theta} - i \bm{\hat \phi})\,, \quad \bm{\hat e_+} = -\frac{1}{\sqrt{2}}(\bm{\hat \theta} + i \bm{\hat \phi}).
\end{equation}
These functions satisfy the relations $\bm{\hat e_-} \cdot \bm{\hat e_-} = \bm{\hat e_+} \cdot \bm{\hat e_+} = 0$ and $\bm{\hat e_-} \cdot \bm{\hat e_+} = \bm{\hat e_+} \cdot \bm{\hat e_-} = -1$. Therefore, we can evaluate the horizontal integral as follows,
\begin{align}
\int &Y_{lm}^* \bPsi_{l'm'} \cdot \bPsi_{lm} \,\id{\Omega} \nonumber \\
&= \int Y_{lm}^* {\Omega_0^{l'}} \,(Y_{l'm'}^- \bm{\hat e_-} + Y_{l'm'}^+ \bm{\hat e_+}) \, {\Omega_0^{l}} \,(Y_{lm}^- \bm{\hat e_-} + Y_{lm}^+ \bm{\hat e_+}) \,\id{\Omega} \notag\\
&= -{\Omega_0^{l}\Omega_0^{l'}} \, \Big(\int Y_{lm}^* Y_{l'm'}^- Y_{lm}^+ \,\id{\Omega} + \int Y_{lm}^* Y_{l'm'}^+ Y_{lm}^- \,\id{\Omega} \Big) .\label{eq:T2angular}
\end{align}
Now, Eq.~\eqref{eq:T2angular} can be written in terms of Wigner 3-j symbols using Gaunt's formula (see Appendix~\ref{appGaunt}),
\begin{align}
\int &Y_{lm}^* \bPsi_{l'm'} \cdot \bPsi_{lm} \,\id{\Omega} \nonumber\\
&=  {4\pi\,(-1)^{m+1} \,\Omega_0^l \Omega_0^{l'} \, \gamma_l \gamma_l \gamma_{l'}} \begin{pmatrix} l & l' & l \\ -m & m' & m \end{pmatrix} \nonumber \\
&\quad\times\left[\begin{pmatrix} l & l' & l \\ 0 & -1 & +1 \end{pmatrix} + \begin{pmatrix} l & l' & l \\ 0 & +1 & -1 \end{pmatrix} \right].\label{eq:angular2}
\end{align}
Since reversing the sign of the lower row in the Wigner 3-j symbol gives a phase factor,
\begin{equation}
\begin{pmatrix} 
l & l' & l \\ 
0 & -1 & +1 \end{pmatrix} 
= (-1)^{2l+l'} 
\begin{pmatrix} 
l & l' & l \\ 
0 & +1 & -1 
\end{pmatrix} \,,
\end{equation}
the integral \eqref{eq:angular2} vanishes if $l'=l_g$ is odd. Consequently, Eq.~\eqref{eq:angular2} results in the same selection rules as Eq.~\eqref{eq:angular1}.

Next, we consider the angular integral appearing in Eq.~\eqref{eq:int2-12}. We now use the relations
\begin{equation}
\bPsi_{lm}^* = {\Omega_0^l} \,(Y_{lm}^{-*} \bm{\hat e^-} + Y_{lm}^{+*} \bm{\hat e^+})
\end{equation}
and $\bm{\hat e^\pm} = \bm{\hat e_\pm^*}$ \citep[see Eq.~C.50 in][]{Dahlen1998GlobalSeismology}, which satisfy $\bm{\hat e^-} \cdot \bm{\hat e_-} = \bm{\hat e^+} \cdot \bm{\hat e_+} = 1$, and $\bm{\hat e^-} \cdot \bm{\hat e_+} = \bm{\hat e^+} \cdot \bm{\hat e_-} = 0$. We then obtain,
\begin{align}
\int& \bPsi_{lm}^* Y_{l'm'} \cdot \bPsi_{lm} \,\id{\Omega} \nonumber \\
&= {\Omega_0^l\Omega_0^l} \, \left(\int Y_{lm}^{-*}  Y_{l'm'} Y_{lm}^- \,\id{\Omega} + \int Y_{lm}^{+*} Y_{l'm'} Y_{lm}^+ \,\id{\Omega} \right)\\
&= {4\pi\,(-1)^{m+1} \,\Omega_0^l \Omega_0^{l} \, \gamma_l \gamma_l \gamma_{l'}}\begin{pmatrix} l & l' & l \\ -m & m' & m \end{pmatrix}\nonumber\\
&\quad\times\left[\begin{pmatrix} l & l' & l \\ 1 & 0 & -1 \end{pmatrix} + \begin{pmatrix} l & l' & l \\ -1 & 0 & 1 \end{pmatrix}  \right], \label{eq:angular3}
\end{align}
which again follows the same selection rules as the above angular integrals.

Similarly, we obtain
\begin{align}
\int  \bPsi_{lm}^* \cdot \bPsi_{l'm'} Y_{lm} \,\id{\Omega}
&= {4\pi\,(-1)^{m+1} \,\Omega_0^l \Omega_0^{l'} \, \gamma_l \gamma_l \gamma_{l'}} \begin{pmatrix} l & l' & l \\ -m & m' & m \end{pmatrix}\nonumber\\
&\quad\times\left[\begin{pmatrix} l & l' & l \\ 1 & -1 & 0 \end{pmatrix} + \begin{pmatrix} l & l' & l \\ -1 & 1 & 0 \end{pmatrix}  \right], \label{eq:angular4}
\end{align}
which again has the same selection rules.

We now turn to the selection rules from the flow terms derived in Appendix~\ref{appFlow}. We begin by noting two common factors in all terms involved in the  integral from the flow term, which therefore can be written (see Eqs.~\ref{eqFlowTermGeneral} - \ref{eqBlllDef})
\begin{align}
\langle \bxi_p, \delta\cL_5 [\bxi_p] \rangle  &= (1+(-1)^{2l+l'}) \begin{pmatrix} l & l' & l \\ -m & m' & m \end{pmatrix} 
\,\times (\text{further terms}).
\end{align}
As a consequence, the contribution from the g-mode induced flow has the same selection rules as the perturbation to solar structure.

\section{Gaunt's formula}
\label{appGaunt}

The surface integral of the product of three generalised spherical harmonics {$Y_{l m}^{N}$} can be written in terms of Wigner 3-j symbols {using the following formula} (see, e.g. \citealp{Dahlen1998GlobalSeismology}, Eqs.~C.188 and~C.198), which holds if $N_1 = N_2 + N_3$,
\begin{align}
\int Y_{l_1 m_1}^{N_1*} Y_{l_2 m_2}^{N_2} Y_{l_3 m_3}^{N_3} \,\id{\Omega} 
&= {4\pi}\, (-1)^{N_1 + m_1} \, {\gamma_{l_1}\gamma_{l_2}\gamma_{l_3}}\nonumber \\ 
&\quad\times\begin{pmatrix} l_1 & l_2 & l_3 \\ -N_1 & N_2 & N_3 \end{pmatrix}
\begin{pmatrix} l_1 & l_2 & l_3 \\ -m_1 & m_2 & m_3 \end{pmatrix} \,.
\label{eq:3sph}
\end{align}
{The definition of generalised spherical harmonics follows slightly different conventions in the literature. We here follow the definition in \citet[][Eqs. C.90, C.117, C.112]{Dahlen1998GlobalSeismology},
\begin{align}
    Y_{lm}^N(\theta,\phi) &= X_{lm}^N(\theta) e^{i m \phi}, \\
    X_{lm}^N(\theta) &= \sqrt{\frac{2l +1}{4\pi}} P_{lm}^N(\cos \theta), \\
    P_{lm}^N(x) &= \frac{1}{2^l} \sqrt{\frac{1}{(l+N)!(l-N)!}} \sqrt{\frac{(l+m)!}{(l-m)!}} \nonumber \\
    &\quad \times (1-x)^{-\frac{1}{2}(m-N)} (1+x)^{-\frac{1}{2}(m+N)} \nonumber \\
    & \quad \times \dv[l-m]{}{x}[(x-1)^{l-N}(x+1)^{l+N}].
\end{align}}
We note that $Y_{lm}^0 = Y_{lm}$.

The Wigner 3-j symbols follow certain selection rules. The second Wigner 3-j symbol in Eq.~\eqref{eq:3sph} is zero unless both of the following conditions are satisfied,
\begin{equation}
\begin{cases}
&m_1 = m_2 + m_3 \, ,\\
&|l_i- l_j| \le l_k, \text{ for } i,j,k = 1,2,3\, .
\end{cases}
\label{eq:sel}
\end{equation}
Additionally, this Wigner 3-j symbols evaluates to zero, if
\begin{equation}
m_1 = m_2 = m_3 = 0 \quad \text{ and } \quad l_1 + l_2 + l_3 \text{  is odd.} 
\label{eq:oddrule}
\end{equation}

\section{Solar models and surface boundary conditions}
\label{appBC}

The numerical results presented in Section~\ref{secResults} depend to some extent on the solar model and surface boundary condition used. This dependence is summarised in Section~\ref{secDependenceOnModelandBC}. In the following, we give some details on solar model and surface boundary conditions that we used.

\subsection{Solar models used}

We use the following solar models, (i) model S \citep{JCD1996}, (ii) model S, where the density was slightly smoothed in radial direction (see details outlined in Appendix~\ref{appSmoothedModelS}), and (iii) a stellar model obtained by evolving a $1M_\sun$ star from pre-main sequence to an age of $4.6\,\rm{Gy}$ using \MESA~\citep{Paxton2011}. The stellar model from \MESA~was obtained using the sample input files in the folder \emph{1M\_pre\_ms\_to\_wd} in the \MESA~test suite. We note that this stellar model was not exactly calibrated to solar values. Our main goal in using this model is to get an impression on how sensitive the numerical computation is to non-negligible differences in solar structure.

\subsection{Details on the smoothed version of model S}
\label{appSmoothedModelS}

In order to evaluate the formulas for the frequency shift numerically, it is necessary to compute second derivatives of solar model quantities such as pressure (e.g. in Eq.~[\ref{eq:int2-11}]) and density (e.g. for the first derivative of the Eulerian density perturbation $\rho_{1,g}$ in Eq.~[\ref{eq:int3-3}]). As the second derivative of the density in model S is not {very} smooth {at some locations in the model interior,} we compute a version of Model S that is smoothed locally, thereby keeping the changes in the overall density structure as small as possible.

The density is smoothed radially with a Gaussian window. The width of the Gaussian varies with the radial location in the solar model with a maximum value of $\sigma_{\rm{max}}=4.87\,\rm{Mm}$ in the core and $\sigma_{\rm{min}}=20\,\rm{km}$ at the outer boundary. Between these values, we vary the width smoothly using a cosine bell,
\begin{align}
\sigma(r) &= \sigma_{\rm{min}} + \frac{\sigma_{\rm{max}}-\sigma_{\rm{min}}}{2} \left( 1 + \cos (\pi \tilde x ^2) \right), \\
\tilde x &= \frac{r}{{R_B}}.
\end{align}
Furthermore, when applying the Gaussian window near the surface boundary, we extend the density with the values it would have in an isothermal atmosphere. This is needed in order to guarantee that the second derivative does not have a kink at the boundary.

\subsection{Surface boundary conditions used}

We compute eigenfunctions using the surface boundary conditions implemented in the \GYRE~code \citep{Townsend2013}. These are a surface boundary matched to an isothermal atmosphere as implemented in \ADIPLS~\citep{JCDadipls}, a surface boundary matched to an isothermal atmosphere in the formulation of \cite{Unno1989}, the atmospheric boundary condition described in \cite{Dziembowski1971}, and a vacuum outer boundary condition (equivalent to a zero Lagrangian pressure perturbation, $\nabla\cdot \bxi=0$).

\end{document}